\documentclass[aps,prx,twocolumn,superscriptaddress,longbibliography]{revtex4-2}
\usepackage{amsmath, amsthm}
\usepackage{keyval}
\usepackage{graphicx,latexsym}
\usepackage{dcolumn}
\usepackage{bm}
\usepackage{color}
\usepackage{ulem} 
\usepackage{hyperref}
\usepackage{xr}

\definecolor{magenta}{rgb}{1,0,1}
\usepackage{color}

\hypersetup{
    colorlinks=true, 
    linktoc=all,   
    linkcolor=blue,  
}

\begin{document}
\title{Dopant-induced stabilization of three-dimensional charge order in cuprates }
\author{Zheting Jin}
\affiliation{Department of Applied Physics, Yale University, New Haven, Connecticut 06520, USA}
\author{Sohrab Ismail-Beigi}
\affiliation{Department of Applied Physics, Yale University, New Haven, Connecticut 06520, USA}
\affiliation{Department of Physics, Yale University, New Haven, Connecticut 06520, USA}
\affiliation{Department of Mechanical Engineering and Materials Science, Yale University, New Haven, Connecticut 06520, USA}
\date{\today}
\begin{abstract}
We investigate the microscopic mechanisms behind the stabilization of three-dimensional (3D) charge order by Pr doping in YBa$_2$Cu$_3$O$_7$ (YBCO7). Density-functional-theory calculations locate the lowest-energy Pr superlattices for both Ba- and Y-site substitution. In the Ba-site case, the smaller Pr ion pulls the surrounding atoms inward. This breathing-mode distortion pins charge-stripe walls to the Pr columns and forces them to align along the $c$ axis. Y-site Pr is larger than the host ion, produces an outward distortion, and fails to pin the stripes. Coarse-grained Monte-Carlo simulations show that the stripe correlation length rises in step with the structural correlation length of the Pr dopant as observed in prior experiments. We thus identify Ba-site substitution and dopant-induced lattice pinning as the key mechanism behind 3D charge order in Pr-doped YBCO7. This approach provides quantitative guidelines for engineering electronic orders through targeted ionic substitution.
\end{abstract}

\maketitle
\section{Introduction}

Charge and spin stripe orders \cite{zhang2002competing,sachdev2003colloquium,vishik2018photoemission} are central to the physics of hole-doped cuprate superconductors. They appear as periodic modulations of charge density and spin orientation within the CuO$_2$ planes and have long been regarded as closely linked with superconductivity \cite{kivelson1998electronic,lee2006doping,chang2012direct,thampy2014rotated,ichikawa2000local,berg2009striped}. 
Stripe order is typically quasi-two-dimensional: correlations are strong in-plane ($a-b$ plane) but lose coherence from layer to layer out-of-plane ($c$ axis). X-ray and neutron scattering measurements find correlation lengths that can remain long in the $a$–$b$ directions yet are orders of magnitude shorter along the $c$ axis in cuprate systems \cite{tranquada1996neutron,comin2016resonant,fujita2011progress,vojta2009lattice,tranquada2020cuprate,j2006magnetic,ghiringhelli2012long,frano2020charge}. This pronounced anisotropy reflects the two-dimensional nature of the Cu $3d_{x^{2}-y^{2}}$ orbitals and the weak interlayer coupling in cuprates\cite{hussey1996classification,jin2025interlayer}.

However, under certain conditions, cuprates can exhibit stable 3D charge-ordered states.  In YBa$_2$Cu$_3$O$_7$ (YBCO7), earlier routes to 3D charge order (CO) involved the application of magnetic fields \cite{gerber2015three, chang2016magnetic, jang2016ideal}, epitaxial strain in thin films \cite{bluschke2018stabilization}, uniaxial strain \cite{kim2018uniaxial, kim2021charge}, or optical pumping \cite{jang2022characterization}. Under these external influences, a relatively sharp 3D CO peak appears at integer L-values (reciprocal lattice points along $c$). However, these 3D CO correlation lengths remain much shorter than typical structural $c$-axis correlation lengths in this compound. More recent work demonstrated that Pr substitution with a certain doping level (Pr$_{x+y}$Y$_{1-x}$Ba$_{2-y}$Cu$_3$O$_{7}$, with $x+y\approx0.3$) can very efficiently stabilize a highly correlated 3D charge-ordered state in YBCO7 with an extremely sharp CO diffraction peak with width limited only by the sharpness of the structural peak \cite{ruiz2022stabilization}. The Pr dopant therefore introduces a yet-unresolved mechanism that ``pins'' the CO phase between adjacent CuO$_2$ planes.

Beyond electronic order, dopants themselves tend to self-organize. In YBCO7, oxygen atoms in the Cu–O chain layer form “Ortho-\textit{n}” phases in which oxygen atoms and vacancies arrange into superstructures with well-defined periodicities \cite{koblischka2021ba2cu3o7}. The Ortho-II phase, for instance, consists of alternating filled and empty Cu–O chains, producing a stripe-like dopant pattern \cite{zimmermann2003oxygen,yamani2004cu,liang2000preparation}. The more complex Ortho-III and Ortho-VIII phases realize distinct oxygen-vacancy patterns \cite{zimmermann2003oxygen,achkar2012distinct}. These ordered dopant superstructures can attain sizable correlation lengths \cite{ricci2013multiscale}. Not limited to the YBCO7 system, similar dopant ordering has also been observed in other cuprate families \cite{song2019visualization, zhang2022visualization, campbell2015long, zhang2022atomic}.

Charge-order and dopant-order symmetry breaking can coexist in cuprates.  In DyBa$_2$Cu$_3$O$_{6+x}$, oxygen-vacancy ordering yields an Ortho-II structural peak at $\mathbf{q}_{\text{Ortho-II}}=(0.5,0,0)$ that is visible at both the Cu $L_{3}$ and Dy $M_{5}$ X-ray edges, whereas the CO peak appears at $\mathbf{q}_{\text{CO}}\approx(1/3,0,0)$ exclusively at the Cu $L_{3}$ edge, allowing the two orders to be cleanly separated \cite{betto2020imprint}.  
The situation becomes less clear when the characteristic wave vectors coincide: in Pr-doped YBa$_2$Cu$_3$O$_7$, resonant soft-x-ray scattering (RSXS) reveals a pronounced three-dimensional peak at $\mathbf{q}\approx(1/3,0,1)$ on the Pr $M_{5}$ edge and a much weaker one with the same wave vector on the Cu $L_{3}$ edge \cite{ruiz2022stabilization}.  The two signals are strongly coupled to each other and share nearly identical temperature dependences. To date, the microscopic origin of this relation remains unclear.

In this work, we tackle the puzzle of why Pr substitution in YBa$_2$Cu$_3$O$_7$ produces an exceptionally sharp, three-dimensional charge-order peak.  Using extensive density functional theory (DFT) simulations, we hypothesize that this charge order is driven by Pr dopants that have substituted on the Ba sites which then order at low temperatures.  To arrive at this hypothesis, we explore the most energetically favorable positions and alignments of the Pr dopants with Ba- and Y-site substitution. The Pr dopants are ordered in both substitution cases and create a superlattice. By contrasting pristine YBCO7 with Ba-site and Y-site Pr-doped variants, we then demonstrate how Pr doping on the Ba-site provides a proper lattice distortion that efficiently enforces the alignment of the 2D CO domain walls along the out-of-plane direction, promoting a stable 3D CO.  In contrast, Pr doping on the Y-site cannot stabilize such an alignment due to improper dopant size. By mapping DFT energetics onto concise classical Hamiltonians for both the Pr sub-lattice and the Cu charge/spin stripes, we perform coarse-grained Monte Carlo simulations to quantitatively estimate the temperature dependency of correlation lengths.  We show that the charge-order correlation length grows in lock-step with the structural correlation length of the Pr arrangement, mirroring RSXS observations \cite{ruiz2022stabilization}.  Our results identify dopant-induced lattice pinning as the dominant mechanism underlying three-dimensional CO in Pr-doped YBCO7 and provide a predictive framework for engineering electronic textures through proper ionic substitution.

\section{Relevant observations and theoretical approach}

Our objective is to understand the nature of the charge order reported in Ref.~\cite{ruiz2022stabilization} using first principles theory.  In that work, it was assumed that Pr substitution occurs on the Y sites, but no experiments were done to check on the atomic site where Pr substitution occurs. assumption.  Indeed, the bulk of the literature on Pr-doped cuprates assumes Y-site substitution assumption.  However, there have been strong suggestions, but not conclusive evidence, of Pr substituting on Ba sites in YBCO7 (e.g., Refs.~\cite{COLONESCU1999528,Kramer1997Suppression,Gasumyants2000EffectofPr}).  Our recent collaborative work~\cite{yang2025rapid} (which includes authors from Ref.~\cite{ruiz2022stabilization}) shows that multiple Pr-doped YBCO7 samples have significant Pr substitution on Ba sites.  These conclusions are based on compositional analysis, x-ray diffraction, and scanning transmission electron microscopy.  And the sample growth approach is the same for both works.  Hence, it is reasonable to consider and compare the consequences of Pr doping on both Y and Ba sites in YBCO7 when explaining the unusual static 3D charge ordering observed.

We now discuss the appropriateness of our theoretical approach for this system and its static charge ordering.  We will be using DFT-based methods in our work: DFT is a method that can describe the ground-state electronic density as well as a material's lowest-energy lattice structure~\cite{hohenberg1964inhomogeneous}.  This is typically done within Kohn-Sham formalism~\cite{kohn1965self}: this is band theory and is a mean-field type effective description of electrons moving independently in a self-consistently determined potential. As such, there is no basis to believe DFT-based outputs for spectroscopy (band widths, gaps, etc.) have much accuracy without direct verification.  And the entire mean-field band structure apparatus should be viewed with skepticism when describing electronic excitations and dynamical processes in strongly correlated materials like cuprate superconductors.  

The subject of charge and spin ordering in cuprate superconductors is a rich and complex field of research.  Many of the interesting and open questions in this field have to do with the dynamical nature of the charge/spin density ordering in space and time, and these questions are beyond the scope of almost all DFT-type predictions.  However, when the spin/charge density ordering becomes static (``freezes out'') and becomes a ground-state property, there is every reason to believe that DFT-type descriptions should be able to describe such static motifs. Literature within the last decade explicitly demonstrates successful DFT-based prediction of the correct ground state structure and basic electron density in multiple 3d transition metal oxides \cite{varignon2019origin} as well as longer-ranged orderings in YBCO7~\cite{zhang2020competing} in Bi$_2$Sr$_2$CaCu$_2$O$_{8+\delta}$~\cite{jin2023afm}.  For these reasons, using DFT-based methods for the narrow task of understanding static ground state structural properties is appropriate and potentially insightful.

\section{DFT-predicted orderings}

Differing dopant arrangements and the associated lattice distortions can yield substantial energy differences in cuprates, usually tens or even hundreds of meV per dopant atom \cite{schwingenschlogl2012doping, jin2024first, varignon2019origin, liu2023first,poloni2018probing, lee2006oxygen}, and these in turn affect the more subtle electronic charge and spin orders in the CuO$_2$ planes. Changing the stripe order domain wall arrangements on the CuO$_2$ planes causes much smaller energy differences, usually below 10 meV per unit cell of domain wall length \cite{jin2024first, zhang2020competing}.  Hence, to organize our search for low-energy states, we begin with the high-energy question of dopant positioning before considering the lower-energy electronic ordering (this important assumption of differing energy scales is checked explicitly below).  

\begin{figure*}[t]
\begin{center}
\includegraphics[scale=0.4]{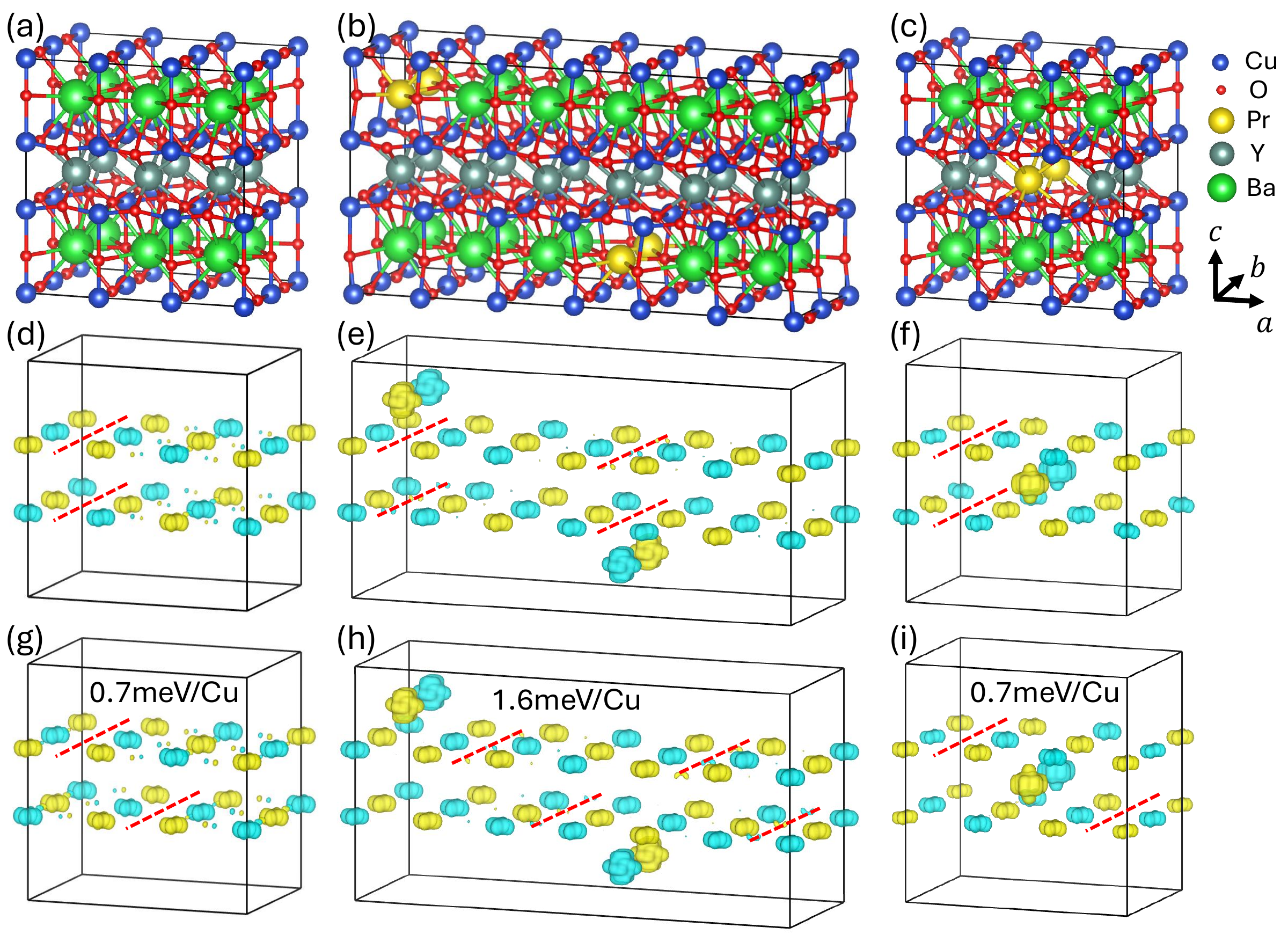}
\end{center}
\caption{  
DFT lowest-energy structure and dopant positions for (a) pure YBCO7, (b) Ba-site Pr-doped YBCO7, when Pr atoms substitute 1/6 of the Ba atoms, and (c) Y-site Pr-doped YBCO7, when Pr atoms substitute 1/3 of the Y atoms. The black boxes mark out the corresponding unit cell. 
(d-f) Spin-density isosurfaces of the ground states in (a-c), where light yellow and blue represent spin up and down separately. The isosurface level is $0.019\mu_B$/\AA$^3$.  The red dashed lines represent the domain walls of the stripe order in CuO$_2$ planes. All domain walls are along the $b$-axis. Different domain walls are stacked/aligned along the $c$-axis and display a 3D CO. 
(g-i) Spin-density isosurfaces of the lowest-energy spin excitations for each case in (d-f) (vertically aligned corresponding figures); corresponding energy increases compared to their ground state energies in (d-f) are listed in units of meV per planar Cu atom.  
}
\label{fig:DFT_charge_order}
\end{figure*}

We first relax a set of structures using DFT calculations where the electronic spin degree of freedom is suppressed, i.e., a constraint of zero magnetization on all atoms.
Following recent experiments \cite{ruiz2022stabilization}, we focus on pure YBCO7 and Pr$_{x+y}$Y$_{1-x}$Ba$_{2-y}$Cu$_3$O$_{7}$, with $x+y=1/3$. Experiments report Pr substitution on both Ba and Y sites, but the samples that exhibit stable three-dimensional (3D) CO belong to the Ba-substituted series \cite{yang2025rapid}. Fig.~\ref{fig:DFT_charge_order}(a)-(c) display the  lowest-energy structures predicted by fully-relaxed DFT calculations (using the magnetic constraint described above) for clean YBCO7, Pr on Y sites, and Pr on Ba sites. These three structures were found by enumerating the set of physically inequivalent dopant configurations in various unit cells for each case and performing relaxations (see Appendix \ref{app:metastable_crystal}). We find that moving any Pr dopant away from its ground-state position incurs a relatively large energy penalty, typically 37-901 meV per Pr ion (as per Appendix \ref{app:metastable_crystal}), an order of magnitude larger than the spin-excitation energies discussed below. 

With the dopant arrangement fixed in its ground state, we permit magnetization to develop on all the atoms and optimize the magnetic texture. The lattice is once again fully relaxed to accommodate the small displacements induced by the formation and arrangement of the local moments. Following established DFT literature protocols for cuprates \cite{jin2024first, zhang2020competing}, we examined competing stripe-ordered and AFM magnetic states containing antiphase domain walls separating antiferromagnetic regions. Both bond-centered (on O sites between neighboring Cu atoms) and site-centered (on Cu sites) domain walls were considered at all symmetry-allowed positions. In pure YBCO7, bond-centered configurations are always energetically favored over site-centered ones, consistent with previous DFT results \cite{zhang2020competing}; this preference persists upon Pr doping, independent of substitution site. Figures~\ref{fig:DFT_charge_order}(d)–(f) present the resulting ground-state spin-density isosurfaces we find for the three structural ground states of Fig.~\ref{fig:DFT_charge_order}(a)–(c). Each displays coupled stripe-like charge and spin modulations, with domain walls indicated by red dashed lines. Across a domain wall, the nearest-neighbor Cu moments align parallel, whereas they remain antiparallel elsewhere (the typical low-energy G-type or checkerboard antiferromagnetic pattern seen in cuprates). The charge distributions are closely coupled to the spin orders and share the same domain walls: Cu hole densities are higher when closer to the domain walls, which is expected in prototypical cuprates \cite{berg2009striped,zheng2017stripe,tranquada2020cuprate,kivelson2003detect} and agrees with earlier DFT results \cite{jin2024first,anisimov2004computation,zhang2020competing}. Interestingly, the domain walls are aligned along the $c$ axis in all three ground states, consistent with the 3D charge-order pattern observed experimentally \cite{ruiz2022stabilization}. In addition, all DFT results indicate a weak antiferromagnetic (AFM) interaction between Pr moments, where we find an in-plane nearest-neighbor AFM exchange coupling of about 1 meV, in line with the $\sim17K$ N\'eel temperature reported experimentally \cite{kebede1989magnetic}.  This small magnetic energy scale does not affect the identification of the lowest-energy magnetic structure on the CuO$_2$ planes.

Although all our DFT calculations indicate an intrinsic preference for charge-order domain walls to align along the $c$ axis, the energetic strength of such a preference is important and can be quantified by computing excitation energies to other lowest-energy metastable states that can disrupt the $c$-axis ordering. Figures~\ref{fig:DFT_charge_order}(g)–(i) display the lowest energy metastable configurations, in which the domain walls are no longer aligned along the $c$-axis. Notably, the energy penalty is small in (g) pristine YBCO7 and (i) Y-site Pr-doped YBCO7 (0.7 meV per Cu), but more than doubles to 1.6 meV per Cu in (h) Ba-site Pr-doped YBCO7. Hence, 3D ordering of the domain walls aligned along the $c$-axis is most energetically stable for Ba-site Pr-doped YBCO7, and in the next subsection, we analyze the microscopic mechanism responsible for this enhanced stability.

\section{Structural pinning effect}
\begin{figure}[t]
\begin{center}
\includegraphics[scale=0.48]{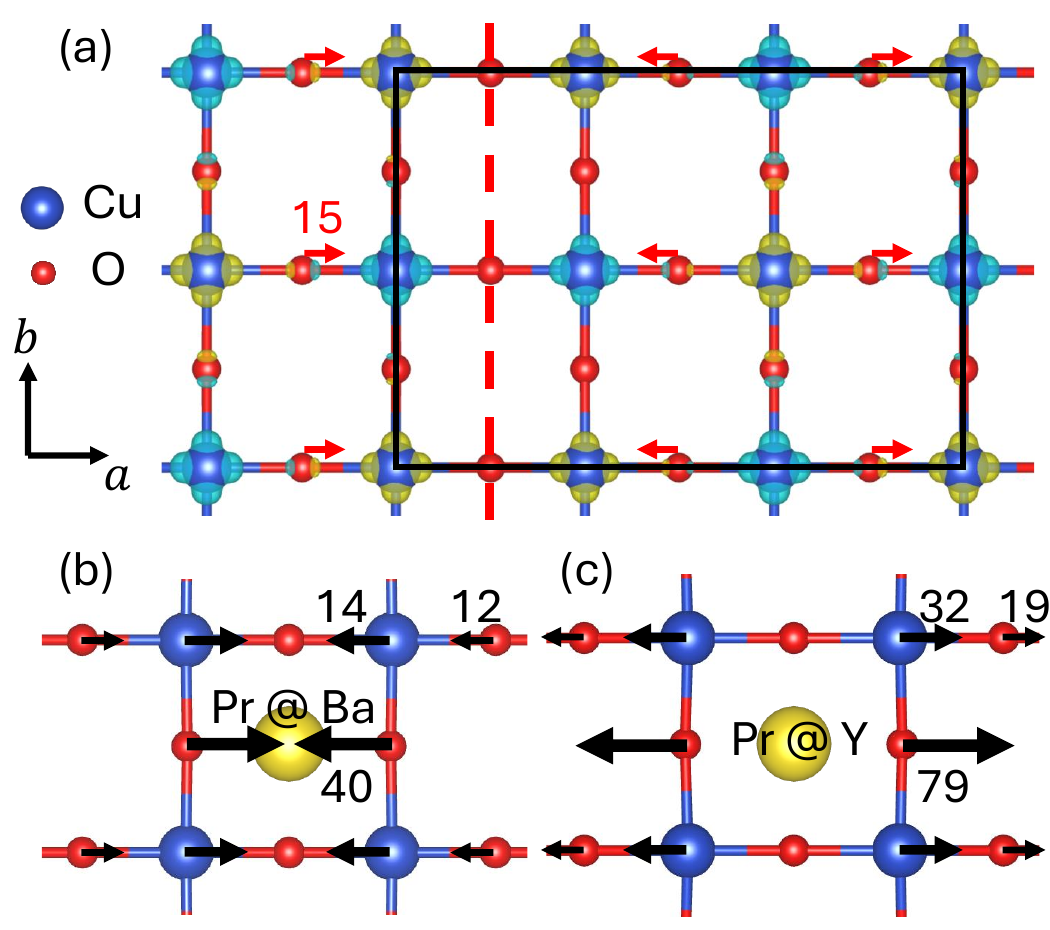}
\end{center}
\caption{
Illustration of local structural distortions due to electronic orders or presence of dopants.  All atomic displacements are in units of $10^{-3}$ \AA.
(a) Top view of one CuO$_2$ plane in pure YBCO7 with stripe order ($q_a=1/3$) as shown in  Fig.~\ref{fig:DFT_charge_order}(d). The isosurfaces shown are of the spin density with yellow and blue marking opposite signs.  The red dashed line marks the electronic domain wall in one unit cell.  The black rectangle is the unit cell of the domain wall pattern.  The red arrows show the atomic displacements of planar oxygen atoms caused by the domain wall. 
(b) Top view of a CuO$_2$ plane when a Pr dopant (yellow ball) replaces the nearest neighbor Ba. The black arrows show the resulting atomic displacements. We have forced the Cu atoms to be non-magnetic to avoid additional electronic ordering effects (e.g., those shown in panel (a)).  (c) The local structural distortion caused by a Pr dopant replacing the Y (with Cu magnetism  suppressed).
}
\label{fig:distortion}
\end{figure}

Stripe ordering and dopant substitution each introduce characteristic lattice distortions in cuprates. Charge stripes, in particular, usually attract/repel atoms to/from their domain walls due to the accumulated charge at the domain wall. For the case of pure YBCO7, Fig.~\ref{fig:distortion}(a) illustrates using red arrows that the dominant deformation is an in-plane shift of the planar oxygen anions (in the CuO$_2$ plane) toward the domain walls by roughly 0.015 \AA, in agreement with earlier DFT predictions \cite{zhang2020competing}. This collective motion constitutes a breathing-mode lattice distortion and was observed using hard-x-ray diffraction \cite{niemoller1999charge,chang2012direct}.
Microscopically, the distortion arises from the strong coupling among spin, charge, and lattice degrees of freedom: doped holes disrupt the AFM background and aggregate along the domain walls, forming hole-rich charge stripes. The positive charge along the domain wall attracts nearby O$^{2-}$ anions, pulling them toward the domain wall to screen the accumulated charge potential, producing the breathing-mode distortion.

The coupling is bidirectional: a pre-existing lattice distortion can template or pin the electronic stripes and domain walls. For example, Ba doping in La$_{2-x}$Ba$_x$CuO$_4$  stabilizes a low-temperature tetragonal octahedral tilt that reduces the in-plane symmetry and pins the stripes in one direction: static charge stripes emerge only when this tilt is present and are suppressed when the tilt is reduced by external pressure~\cite{fabbris2013local}. Separately, in Bi$_2$Sr$_2$CaCu$_2$O$_{8+x}$, the stripe order domain walls prefer to stay away from the oxygen dopants to minimize the total energy \cite{jin2024first}. 

Next, we consider Pr dopants in YBCO7.  For Ba site Pr doping, neighboring ions relax toward the Pr ion as shown in Fig.~\ref{fig:distortion}(b), producing a lattice displacement much larger in magnitude than those generated by the electronic stripes themselves.  The inward motion is easy to understand and is due to Pr$^{3+}$ being smaller than Ba$^{2+}$ \cite{Shannon:a12967}.  If the electronic domain wall were to center itself on the dopant, the preferred atomic motions generated by both would be in the same direction and would reinforce each other to lower the total energy: consequently, we find that charge stripes lock to this template and the domain walls are pinned at the dopant positions, as illustrated in Fig.~\ref{fig:DFT_charge_order}(e). Any attempt to shift a domain wall within a CuO$_2$ layer must overcome the additional dopant pinning energy, consistent with the larger energy cost reported in Fig.~\ref{fig:DFT_charge_order}(h).  

By contrast, Pr substitution at the Y site pushes the surrounding lattice outward, as shown in Fig.~\ref{fig:distortion}(c), because Pr$^{3+}$ is larger than Y$^{3+}$~\cite{Shannon:a12967}. When the domain wall is centered on the dopant, these dopant-induced atomic displacements oppose those favored by the domain wall. As a result, electronic domain walls tend to avoid Y-substituted dopant sites (see Fig.~\ref{fig:DFT_charge_order}(f)). Once a domain wall has bypassed the dopants and relocated to the ``clean'' regions of the CuO$_2$ layer, displacing it in one CuO$_2$ layer relative to the adjacent one (see Fig.~\ref{fig:DFT_charge_order}(i)) incurs only a small energy cost that is comparable to that in pure YBCO7 (compare Figs.~\ref{fig:DFT_charge_order}(g) and (i)).

We point out the implications of our findings on charge ordering (CO) in Pr doped YBCO.  Fig.~\ref{fig:distortion} clarifies that the formation of electronic domain walls and/or the introduction of Pr dopants create static local structural distortions (i.e., atomic displacements) and necessarily create inhomogeneity in the spatial charge distribution.  When these dopants and/or domain walls order periodically in space, as they do in the lowest energy predicted structures, we necessarily find long-ranged and static 3D CO (and spin ordering).  

In brief, Ba-site Pr dopants should stabilize three-dimensional CO by pinning the domain walls in every CuO$_2$ layer to the dopant positions. Thus far, however, these qualitative conclusions rest on simple comparisons of excitation energies obtained from DFT and are valid at zero temperature. To connect more directly with experiments, we will simulate the temperature-dependent charge-order correlation length.

\begin{figure}[t]
\begin{center}
\includegraphics[scale=0.31]{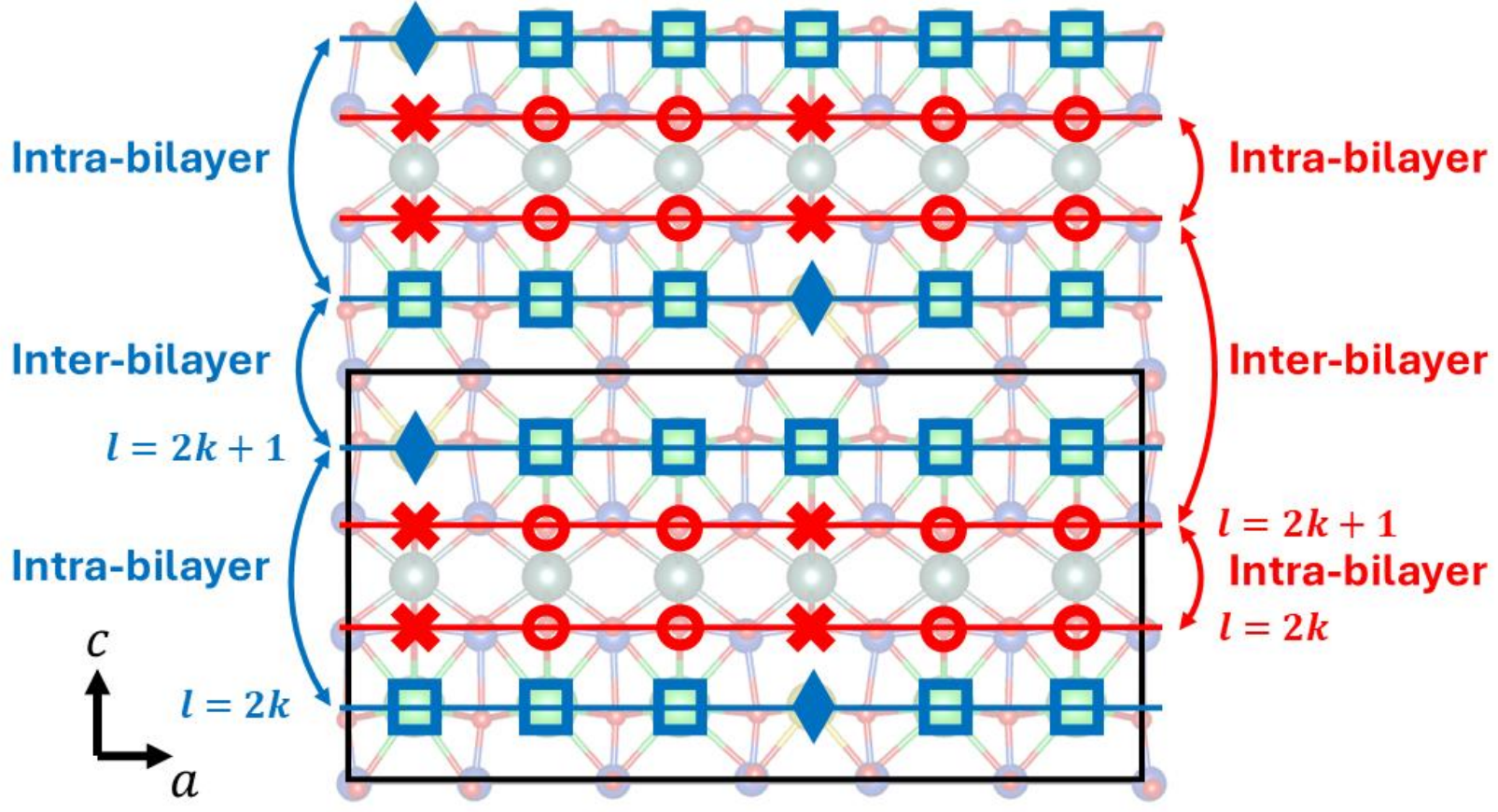}
\end{center}
\caption{
Illustration of a lattice configuration with a dopant and CO domain wall grids in the $ac$ plane. The crystal structure in the background is a side view of Fig.~\ref{fig:DFT_charge_order}(b). Pr dopants (on Ba sites) are shown as blue diamonds while Ba atoms are marked as blue squares. The CO domain wall locations in Fig.~\ref{fig:DFT_charge_order}(e) are marked by red crosses, and red circles represent other possible CO domain wall positions (domain walls are assumed straight in the $b$ direction). The black rectangle represents the unit cell in Fig.~\ref{fig:DFT_charge_order}(b) containing one CuO$_2$ bilayer. We use even layer number $l$ to represent a lower layer in a unit cell, and odd $l$ to represent an upper layer (for dopants or domain walls).  The interlayer interaction in the unit cell is denoted as ``intra-bilayer'', while the interaction across different unit cells is denoted as ``inter-bilayer''. 
}
\label{fig:lattice}
\end{figure}

\section{Correlation lengths}

Our simulations will focus on the experimental doping level and assume Ba site Pr doping: we substitute 1/6 of the Ba atoms in YBCO7 with Pr dopants. This is the only doping level exhibiting stable 3D CO in experiment \cite{ruiz2022stabilization}. For computational efficiency, we model the domain walls as straight and oriented along the $b$ axis (with AFM spin ordering along $b$), parallel to the Cu-O chain direction. (The assumption of straight domain walls along the $b$ axis is explicitly tested and verified in Appendix \ref{app:metastable_mag}
.) With this setup, a domain-wall configuration is fully specified by the wall positions in the $ac$ plane. This assumption is also consistent with prior experimental observations \cite{ruiz2022stabilization, gerber2015three, chang2016magnetic, jang2016ideal} and theoretical studies \cite{zhang2020competing}.  Similarly, we assume dopants form straight linear chains along the $b$ direction.  These assumptions are based on our findings that the energy cost to misalign dopants or domain walls along the $b$ is very large, as shown in Appendices~\ref{app:metastable_crystal} and \ref{app:metastable_mag}.  This makes disordering along the $b$ direction out of the energy window of practical interest for the temperature ranges we will simulate. Fig.~\ref{fig:lattice} illustrates an example of the positions of dopants (blue) and domain walls (red) in the $ac$ plane, where the crystal and spin structures in Fig.~\ref{fig:DFT_charge_order}(b) and (e) are used to build the example.  This representation and associated lattice grid for dopant and domain wall positions underlie a lattice Hamiltonian we construct below.

\begin{table}
\begin{tabular}{c|ccccc}
Inplane distance $\Delta a$ & 0 & 1 & 2 & 3 & $\geq 4$ \\
\hline
$\Delta c=0$ (in plane) &   & 380 & 115 & 15 & 0 \\
$\Delta c=$ intrabilayer NNID & 88 & 87 & 15 & 0 & 0 \\
$\Delta c=$ interbilayer NNID & 166 & 162 & 28 & 0 & 0 \\
\end{tabular}
\caption{DFT-predicted effective interaction $V^{\text{dopant}}_{ili'l'}$ in meV between two Pr dopants separated by $\Delta a \equiv |i-i'|, \Delta c \equiv |l-l'|$ along $a,c$ axes. The in-plane separations $\Delta a$ are in units of the nearest-neighbor Ba-Ba distance (equal to the Cu-Cu distance in the CuO$_2$ plane). For out-of-plane separations $\Delta c$, only the Pr-Pr interactions in plane and between nearest-neighbor layers along the $c$-axis are considered, i.e., $\Delta c=0$ (in plane), $\Delta c=$ intra-bilayer nearest-neighbor interlayer distance (NNID), or $\Delta c=$ inter-bilayer NNID. }
\label{tab:interaction}
\end{table}

We begin with the higher-energy scales involving the dopant configurations and then consider the domain wall energetics.  We start by quantifying the pairwise interaction between Ba-site Pr dopants based on our DFT calculations. To remove the effect of the Cu local moments (which create their own energetics due to domain wall physics, which we deal with separately), all Cu atoms are constrained to be non-magnetic for these computations. Comparing the total energies of ten representative (meta)stable configurations (see Appendix~\ref{app:metastable_crystal}) with that of the ground-state arrangement of Fig.~\ref{fig:DFT_charge_order}(b) allows us to build a simple lattice model for the Ba-substituted Pr dopant interaction energies
\begin{equation}
    E^{\text{dopant}} =\tfrac12 \sum_{i,i',l,l'} V^{\text{dopant}}_{ili'l'}n_{il}n_{i'l'}\,.
\end{equation}
Ba lattice sites in the $ac$ plane are indexed by the integer pairs $(i,l)$ and $(i',l')$ where $i,i'$ label the $a$-axis positions and $l,l'$ label the $c$-axis positions, highlighted in blue in Fig.~\ref{fig:lattice}.  The site $(i,l)$ has dopant occupancy $n_{il}\in \{0,1\}$, and $V^{\text{dopant}}_{ili'l'}$ is the interaction energy between occupied sites $il$ and $i'l'$. We fit ten different DFT energies to such a model containing nine non-zero interaction strengths. Concretely, for each DFT configuration $\alpha$, we compute the energy difference $\Delta E_\alpha \equiv E^{\text{DFT}}_\alpha - E^{\text{DFT}}_{\rm GS}$ relative to the ground state and require the model to reproduce this DFT energy difference.  The resulting interaction strengths are listed in Table~\ref{tab:interaction} (Appendix~\ref{app:metastable_crystal} provides detailed examples showing how specific $\Delta E_\alpha$ values map onto the individual entries of Table~\ref{tab:interaction}.) The DFT-extracted $V^{\text{dopant}}$ couplings turn out to be purely repulsive, decay rapidly with separation, and are assumed to be zero beyond the tabulated range. Note that there are two inequivalent types of interlayer interactions for dopants as illustrated in Fig.~\ref{fig:lattice}. 

\begin{table}
\begin{tabular}{c|cccc}
Inplane distance $\Delta a$ & 0 & 1 & 2 & $\geq 3$ \\
\hline
$\Delta c=0$ (Inplane) &   & 34 & 10 & 0 \\
$\Delta c=$ intrabilayer NNID & -3.0 & 0 & 0 & 0 \\
$\Delta c=$ interbilayer NNID & -1.5 & 0 & 0 & 0 \\
\end{tabular}
\caption{DFT-predicted effective interaction $V^{\text{DW}}_{ili'l'}$ in meV between two CO domain walls separated by $\Delta a \equiv |i-i'|, \Delta c \equiv |l-l'|$ along $a,c$ axes, in unit of nearest Cu-Cu spacing. For the out-of-plane direction, only the interactions in-plane and between nearest-neighbor layers are considered, similar to Table~\ref{tab:interaction}. }
\label{tab:dw_interaction}
\end{table}

Next, we consider the electronic domain wall energetics. We now allow the Cu sites to carry static magnetic moments in the DFT calculations. For both pristine and Pr-doped YBCO7, we identify, in addition to the global minimum, several metastable stripe-ordered states with distinct domain-wall arrangements. Their geometries and energies are compiled in Appendix~\ref{app:metastable_mag}.  
Similarly, by computing the energy differences among these states, we find a lattice model for the interaction energies of the stripe domain walls  
\begin{equation}
    E_{\text{DW}} = \tfrac12\sum_{i,i',l,l'} V^{\text{DW}}_{ili'l'}\tilde n_{il} \tilde n_{i'l'} + \sum_{i,l}\epsilon_{il}\tilde n_{il}\,.
\end{equation}
The lattice sites $(i,l)$ denote the location of the stripe order domain walls in the $ac$ plane, highlighted in red in Fig.~\ref{fig:lattice}.  The domain wall occupancy of site $(i,l)$ is $\tilde n_{i l}\in\{0,1\}$. The first term $V^{\text{DW}}_{ili'l'}$ describes the interaction energy between two domain walls located at $(i,l)$ and $(i',l')$ as tabulated in Table~\ref{tab:dw_interaction}.  We assume that interactions beyond the tabulated range are zero (they are weak and also computationally expensive to compute).  We notice that strong in-plane repulsions keep the domain walls spaced out in the $a$ direction while weak attractions align them in the $c$ direction. The second on-site term $\epsilon_{i l}$ captures the important dopant pinning effect: $\epsilon_{i l}=-4.5\ \mathrm{meV}$ if the nearest Ba site (directly above or below) is substituted by Pr (i.e. $n_{i l}=1$), while $\epsilon_{i l}=0$ otherwise. 

\begin{figure}[t]
\begin{center}
\includegraphics[scale=0.55]{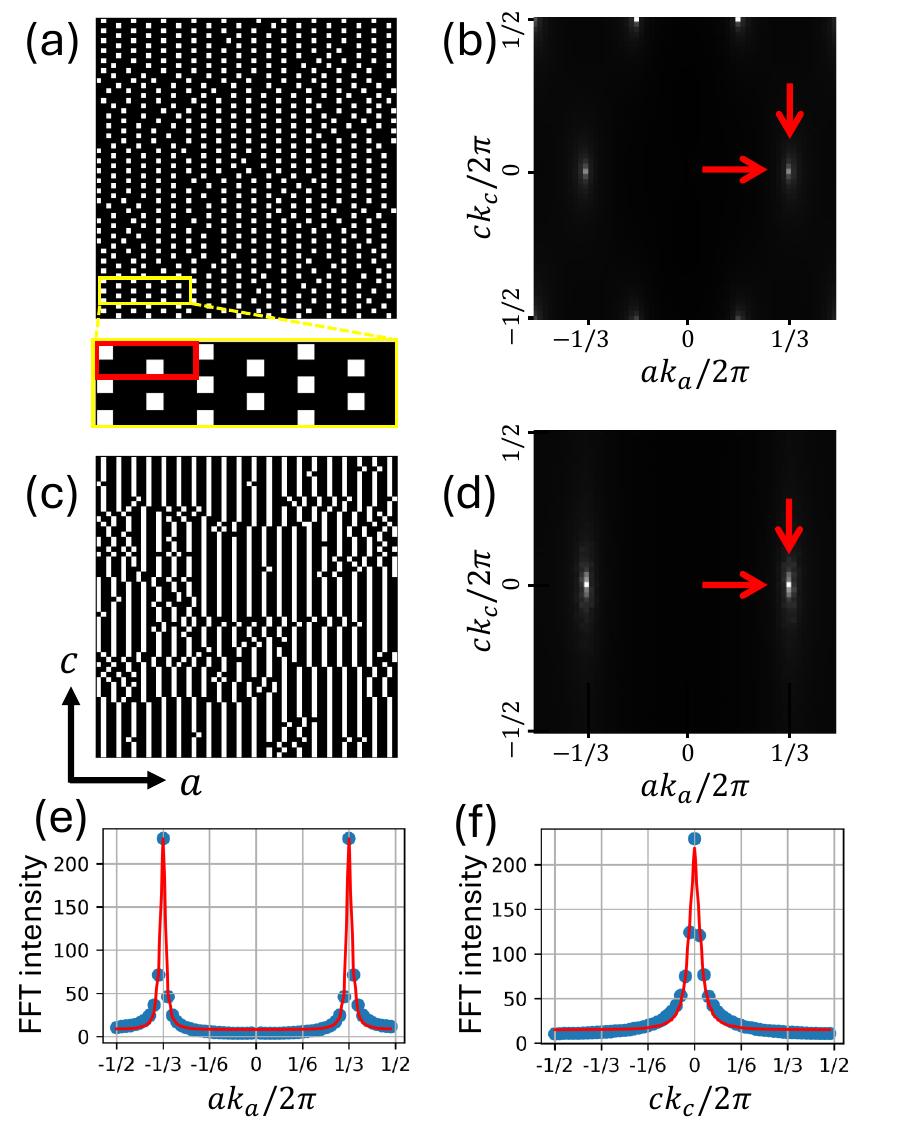}
\end{center}
\caption{
Monte Carlo results for the dopant and domain walls in a $60\times60$ supercell. (a) A snapshot of the dopant positions in real space in the $ac$ plane at 136K, where white ($n_{il}=1$) and black ($n_{il}=0$) pixels represent Pr dopants and Ba atoms, respectively. The lower panel shows a zoomed-in view inside the yellow box, where the red rectangular marks out one DFT unit cell shown in Fig.~\ref{fig:DFT_charge_order}(b). (b) The averaged squared magnitude of the Fourier transform of the dopant configurations averaged over 1,000 snapshots. (c) Snapshot of the CO domain wall positions in real space in the $ac$ plane at 50K, where white/black pixel represents a position with/without a CO domain wall ($\tilde n_{il}$ is 1/0). The dopant alignment in (a) is used for this simulation. (d) The averaged squared magnitude of the Fourier transform of the $\tilde n_{il}$ following the recipe in (b). (e) and (f): The peak intensities along the in-plane and out-of-plane direction marked by the red arrows in (d).  Blue dots show the computed Monte Carlo data, while the red curves are Lorentzian fits using a simple least-squares fit.  
}
\label{fig:monte_carlo}
\end{figure}

We now use the above lattice models and energy functions to compute temperature-dependent spatial dopant correlation functions using classical statistical mechanics via canonical Monte Carlo sampling.  While the use of classical statistical mechanics for predicting thermalized dopant configurations should be quite safe (Pr is a heavy atom and likely faces large eV-scale energy barriers to move between Ba sites), for the domain walls this ignores, at a minimum, their quantum tunneling between nearby sites as well as the associated quantum fluctuations.  However, a full quantum statistical treatment of the domain wall physics is very difficult and beyond the scope of this initial work.  The only quantum feature we consider regarding the domain walls is that their dynamics must be {\it fast} compared to the dopants: namely, as per the Born-Oppenheimer approximation, we assume for each fixed dopant configuration, the domain walls can quickly thermalize as appropriate for that dopant layout.  Hence, our approach will only be able to make predictions about static distributions of dopants and domain walls, but this is enough for our target of describing the static CO observed in experiments.  

Since the domain-wall energetics are coupled to the dopant configuration via the pinning term but not the other way around, we simulate the coupled system by first finding the thermal distribution of dopants, and then for each fixed dopant configuration drawn from a thermal distribution, we find the thermal distribution of the domain walls. 

We use a hybrid Monte Carlo scheme that combines Metropolis–Hastings updates \cite{metropolis1953equation} and Swendsen–Wang cluster moves \cite{swendsen1987nonuniversal}.
During a Metropolis step, a single dopant or CO domain wall is proposed to hop to a randomly chosen nearest-neighbor site.  During a Swendsen–Wang step, a randomly selected collection of random number of adjacent rows is shifted horizontally by 1, 2, or 3 lattice sites. Both update types conserve the total number of particles (or CO domain walls), as required. Our hybrid algorithm attempts one Swendsen–Wang move after every 40 Metropolis trials. A proposed move of either flavor is accepted with the standard Metropolis probability  $  p=\min \bigl[1,\exp(-\Delta E/k_{\mathrm B}T)\bigr],
$ where $\Delta E$ is the associated energy change. 
To accelerate convergence at low temperatures, we employ parallel tempering \cite{swendsen1986replica,marinari1992simulated}: every 400 Monte Carlo trial steps, we propose an exchange of configurations between neighboring temperatures $T_i$ and $T_j$. The exchange is accepted according to the Metropolis–Hastings criterion with probability
$$
p = \min\left[1,
      \exp\Bigl((E_i - E_j)\Bigl(\tfrac{1}{kT_i} - \tfrac{1}{kT_j}\Bigr)\Bigr)
    \right],
$$
where $E_i$ and $E_j$ denote the total energies of the current configurations in the Markov chains at temperatures $T_i$ and $T_j$, respectively.
We use 12 temperatures for the dopant calculations and 7 temperatures for the domain walls. As detailed in Appendix~\ref{app:mc_details}, this setup ensures that simulations at adjacent temperatures have a reasonable probability of visiting the same states. 
The main results shown below use a $60\times60$ supercell in the $ac$ plane with periodic boundary conditions. We carry out $1.5\times10^{7}$ Monte Carlo iterations for each specific temperature (either for sampling the dopant distribution or for sampling the domain wall distribution at fixed dopant configuration). We find that thermal equilibrium is reached after the first $1.0\times10^{7}$ steps at all temperatures we studied (equilibration is determined via the energy autocorrelation function as per Appendix~\ref{app:mc_details}).  Hence, all statistical data are computed using the post-thermalized remaining $0.5\times10^{7}$ steps.

Before presenting our main results on correlation lengths, we outline the process leading to the extraction of correlation lengths with a representative example. Figure~\ref{fig:monte_carlo}(a) shows a Monte-Carlo snapshot of the dopant occupancy configuration $n_{il}$ on a $60\times60$ supercell at $T = 136$K recorded after equilibration. The yellow rectangle enlarges a region whose local ordering reproduces the DFT ground state of Fig.~\ref{fig:DFT_charge_order}(b): one DFT unit cell corresponds to the red rectangle in the magnified view. The snapshot shows that several such ordered regions are separated by disordered boundaries.
Fig.~\ref{fig:monte_carlo}(b) shows the averaged square magnitude of the 2D fast-Fourier-transform (FFT) of the $n_{il}$ computed from 1,000 snapshots uniformly spaced over the final $5\times10^{6}$ Monte-Carlo steps. Using the dopant configuration in Fig.~\ref{fig:monte_carlo}(a), Fig.~\ref{fig:monte_carlo}(c-d) are the CO domain wall results at 50K presented in the same format as Fig.~\ref{fig:monte_carlo}(a-b). 
Correlation lengths are obtained by fitting the principal FFT peak (red arrows in Fig.~\ref{fig:monte_carlo}) with a Lorentzian profile along both the in-plane and out-of-plane reciprocal-space directions, as illustrated in Fig.~\ref{fig:monte_carlo}(e)–(f) for the peak of panel (b). For the in-plane fit along $k_a$ we set $k_{c}=0$; for the out-of-plane fit along $k_a$ we set $k_{a}=p$, where $p$ is the coordinate with maximum height along $k_a$. The correlation length along a given direction is defined as the inverse of the half-width at half-maximum extracted from the corresponding Lorentzian.

\begin{figure}[t]
\begin{center}
\includegraphics[scale=0.45]{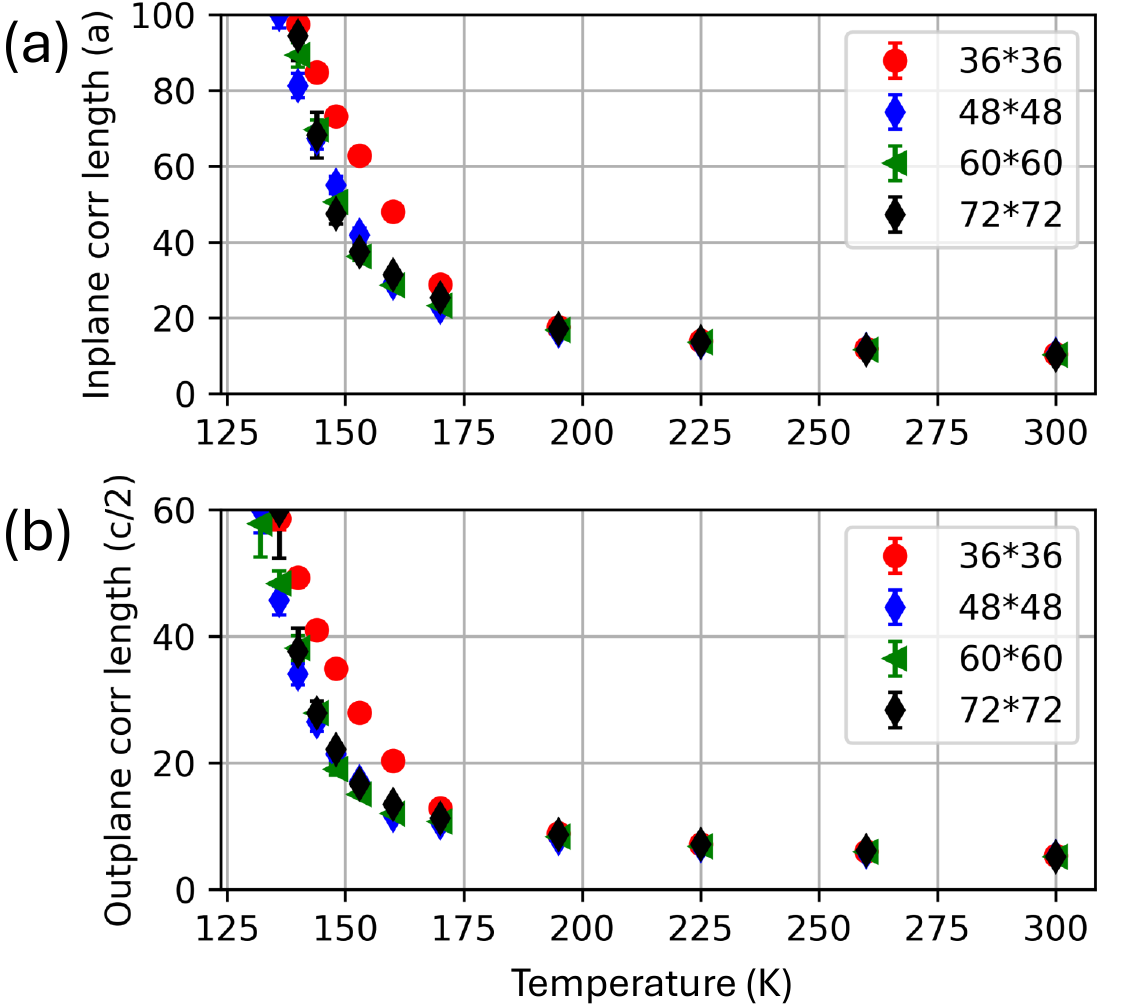}
\end{center}
\caption{
Convergence of correlation lengths with respect to supercell size.
(a) Dopant in-plane (IP) correlation length (CL) as a function of temperature, where the unit of length is the primitive cell lattice constant $a\approx3.8$\AA. Different symbols represent results from different $L\times L$ supercell sizes marked as $L*L$. The error bars of the dopant CLs are the standard deviation of the CLs based on 100 independent MC simulations with different random initial guesses. (b) Dopant out-of-plane (OOP) CL, presented in the same manner as (a). The length unit is half of the primitive cell lattice constant $c/2\approx5.9$\AA.
}
\label{fig:system_size}
\end{figure}

As explained above, the inter-dopant interactions are roughly an order of magnitude stronger than those of the domain walls, so we simulate the dopant subsystem first, ignoring the domain wall stripes. To test convergence versus finite size effects, we run Monte Carlo simulations on several supercells, and  Figure~\ref{fig:system_size} compiles the results. For each supercell size, we perform 100 independent Monte Carlo simulations starting from a random configuration, and we plot the mean correlation length together with its standard deviation. For both in-plane and out-of-plane directions, the correlation lengths obtained from the $60\times60$ supercell are converged versus supercell size for temperatures $T\gtrsim 130\;\text{K}$. Separately, as the temperature is lowered, the correlation lengths grow very rapidly, as shown in Fig.~\ref{fig:system_size}, and they quickly reach the size of the supercell, which makes quantitative calculations at lower temperatures unreliable.  Luckily, since the largest out-of-plane structural correlation length observed experimentally is $\sim 364\;\text{\AA}$ \cite{ruiz2022stabilization}, equivalent to 60 lattice units in our notation, our results with $60\times60$ supercells are sufficient to describe the experimentally relevant regime.

\begin{figure}[t]
\begin{center}
\includegraphics[scale=0.55]{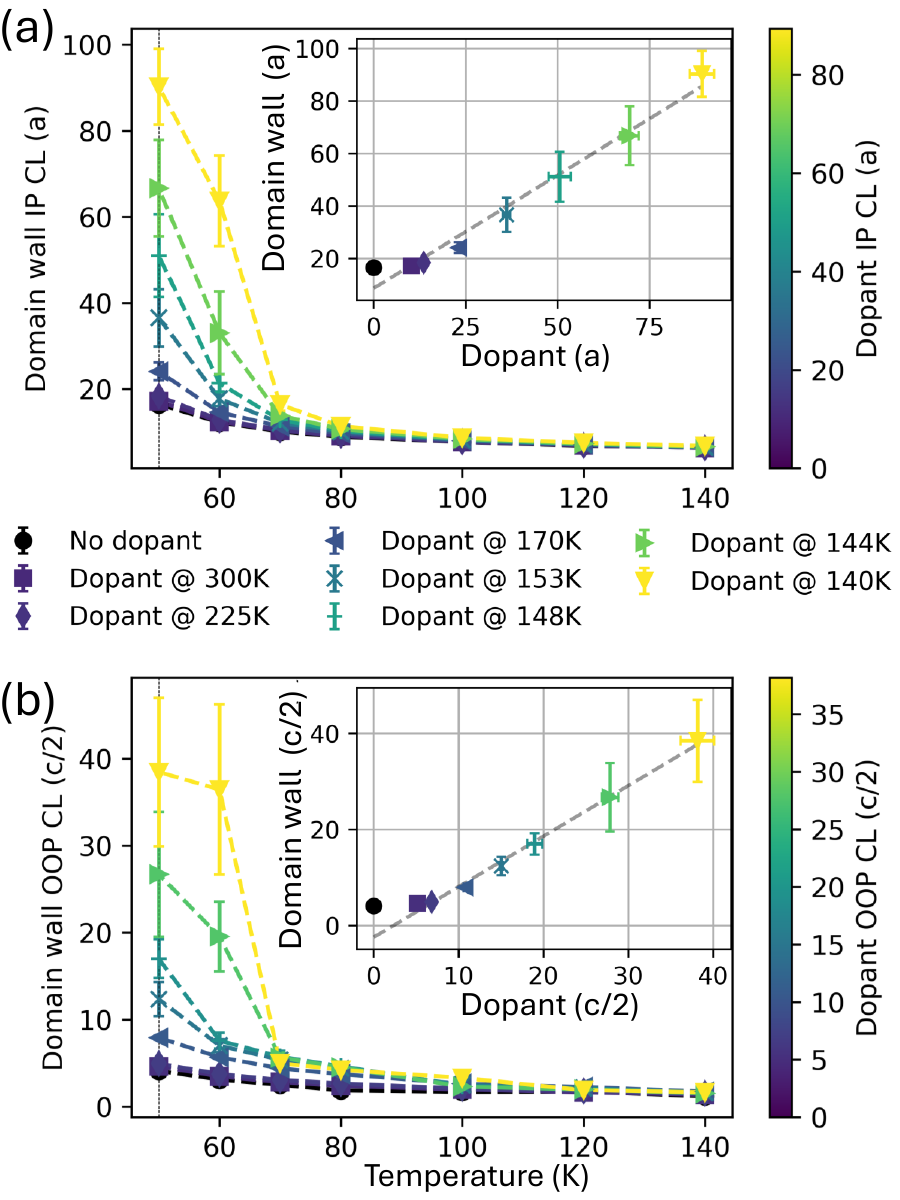}
\end{center}
\caption{
(a) Domain wall IP CL as a function of temperature, simulated in a $60\times60$ supercell. Each colored curve shows the domain wall CL using dopant distributions generated from a fixed temperature, with the corresponding dopant correlation length indicated by the color bar on the right. The inset displays the relationship between the domain wall IP CL at 50~K and the corresponding dopant IP CL, along with a least-squares fit (dashed line). Error bars for the dopant CLs are obtained from 100 independent MC simulations (see Fig.~\ref{fig:system_size}). For each dopant configuration, 10 independent CO domain wall simulations are performed, resulting in a total of 1,000 CO domain wall simulations. The standard deviation across these simulations is shown as the error bar for the domain wall CL at each data point.  
(b) Domain wall and dopant OOP CLs, presented in the same manner as in (a).
}
\label{fig:corr_len}
\end{figure}

Experimentally, the measured structural correlation length at $T = 50\;\text{K}$ is not enormous or divergent \cite{ruiz2022stabilization}, implying that the Pr dopant distribution at 50~K is in fact not at thermodynamic equilibrium and is instead ``frozen''.  In more detail, we start at high temperatures where the dopants diffuse rapidly and their distribution can quickly reach equilibrium.  As we lower the temperature, dopant diffusion rapidly slows down since it is thermally activated, and the experimental cooling rate will eventually outpace the time for equilibration (see Appendix~\ref{app:diffusion} for numerical estimates based on known experimental diffusion barrier). Below some temperature $T_{\mathrm{freeze}}$, the dopant configuration behaves like a quenched snapshot of a higher-temperature equilibrium state. Although, to the best of our knowledge, the diffusion coefficient of Pr in YBCO7 has not been measured directly, it is reasonable to expect the Pr dopants to freeze somewhere between room temperature (300 K) and 50 K during the experimental cool-down \cite{ruiz2022stabilization}. 

Nevertheless, since we lack an exact $T_{\mathrm{freeze}}$, we work around this issue by examining dopant configurations that are equilibrated at several temperatures along the cooling path and study how the resulting dopant correlation length influences the CO domain-wall correlation length.  To this end, we select seven representative temperatures from the dopant simulations with the $60\times60$ supercell (Fig.~\ref{fig:system_size}). For each temperature, we perform 100 independent Monte Carlo runs, each run yielding one thermally equilibrated snapshot of the dopant configuration. Each snapshot then serves as the fixed background for ten independent, parallel-tempered Monte Carlo simulations of the CO domain walls. Figure~\ref{fig:corr_len} displays the resulting in-plane and out-of-plane domain-wall correlation lengths, obtained by averaging over this ensemble. Results for a pristine lattice with no dopants are included for comparison.

The insets of Fig.\,\ref{fig:corr_len} display the behavior at $T = 50\;\text{K}$ (vertical dashed line) and show the relation between the domain-wall correlation lengths and the corresponding dopant correlation lengths. In the absence of dopants, or at high temperatures such as 300 K,  thermal fluctuations dominate and the domain walls remain essentially unpinned, exhibiting a modest correlation length of $\sim14a$ in-plane and $\sim2c$ out-of-plane. Once the dopants begin to order, however, the domain walls lock onto the dopant template, and the CO domain wall correlation length simply follows the dopant correlation length (as shown by the linear fits in the insets). This pinning behavior or mechanism explains why Ba-site Pr substitution so effectively stabilizes 3D CO in YBCO7: experimentally, the CO correlation length grows with cooling until it is limited only by the intrinsic crystalline coherence length \cite{ruiz2022stabilization}.

\section{Summary}
In summary, our combined DFT and large-scale Monte Carlo study reveals that Ba-site Pr substitution in YBa$_2$Cu$_3$O$_7$ stabilizes three-dimensional charge order through lattice distortions that pin stripe domain walls to the dopant positions. The resulting ground state is protected by an energy barrier of about 4 meV per Cu, an order of magnitude larger than in pristine or Y-site-substituted samples.  Lattice models extracted from {\it ab initio} total-energy differences combined with finite temperature classical Monte Carlo sampling reproduce the experimentally observed hierarchy of correlation lengths: as soon as the Pr sub-lattice develops finite structural order, the electronic stripe (domain wall) correlation lengths grow correspondingly and are limited only by the structural coherence length of the dopants. These findings establish static dopant order as a key stabilizer of electronic order in cuprates.  In addition, our results show how ionic size mismatch can be used as a structural design principle to engineer 2D and 3D electronic correlations in cuprates and perhaps other layered oxides.

\section{Methods}

All DFT calculations in this work were performed with the Vienna ab initio simulation package (VASP) software \cite{kresse1996efficiency, kresse1996efficient}. The generalized-gradient-approximation (GGA) with the semilocal Perdew–Burke–Ernzerhof (PBE) functional \cite{anisimov1991band, perdew1996generalized} is used in all of our calculations. To mitigate self-interaction errors~\cite{perdew1981self}, DFT+$U$ corrections was introduced with $U_{\text{Cu}} = 4$ eV applied to the Cu $d$ orbitals and $U_{\text{Pr}} = 6$ eV applied to the Pr $f$ orbitals. The value of $U_{\text{Cu}}$ follows previous theoretical studies~\cite{wang2006oxidation, yelpo2021electronic, deng2019higher, jin2024first, jin2025interlayer} and accurately reproduces the experimental magnetic moments and exchange interactions discussed in Appendix~\ref{app:dftu}. Reported $U_{\text{Pr}}$ values range from 6 eV to 10 eV in earlier work \cite{liechtenstein1995quantitative, ghanbarian2006effects, ghanbarian2008different}, our choice of 6 eV is the most commonly used value. Appendix~\ref{app:dftu} further shows that varying $U_{\text{Pr}}$ between 6 eV and 10 eV has a negligible impact on the energy models presented in the main text and therefore only a minor influence on the principal results. A relatively high plane-wave cutoff energy of 500 eV was used for all calculations. Brillouin-zone sampling employed a $4\times6\times2$ Monkhorst-Pack grid for pristine YBCO7 [Fig.~\ref{fig:DFT_charge_order}(a)] and for Pr substitution at the Y site [Fig.~\ref{fig:DFT_charge_order}(c)]. For Pr substitution at the Ba site [Fig.~\ref{fig:DFT_charge_order}(b)], a $2\times6\times2$ grid was used because the lattice constant doubles along the $a$ axis.

\section{Acknowledgment}

We thank Yu He, Jinming Yang, M. Brian Maple, Alex Frano, and Wei-Wei Xie for their helpful discussions. This work was supported by grant NSF DMR-2237469, NSF ACCESS supercomputing resources via allocation TG-MCA08X007, and computing resources from Yale Center for Research Computing. S.I. thankfully acknowledges partial support from NSF DGE-2244310.


\appendix
\begin{figure}[t]
\begin{center}
\includegraphics[scale=0.5]{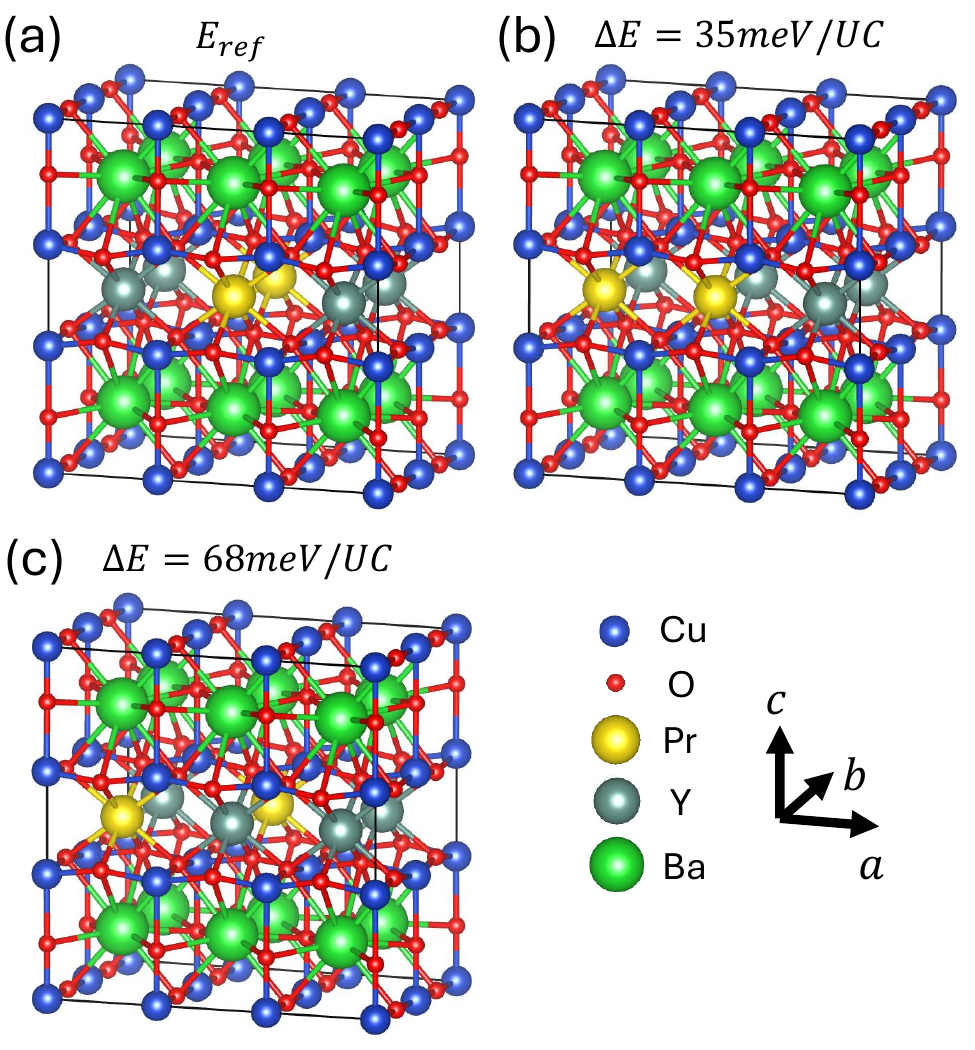}
\end{center}
\caption{  
A set of (meta)stable crystal structures on a $3\times2\times1$ supercell in which $1/3$ of the Y sites are substituted by Pr, corresponding to two Pr atoms per unit cell (outlined by the black box). 
(a) Ground-state structure with the lowest total energy. (b-c) Energy penalties $\Delta E$ are reported relative to the reference energy $E_{ref}$ of structure (a) in the unit of meV per unit cell. 
}
\label{fig:app_Y_2Pr}
\end{figure}
\begin{figure*}[t]
\begin{center}
\includegraphics[scale=0.53]{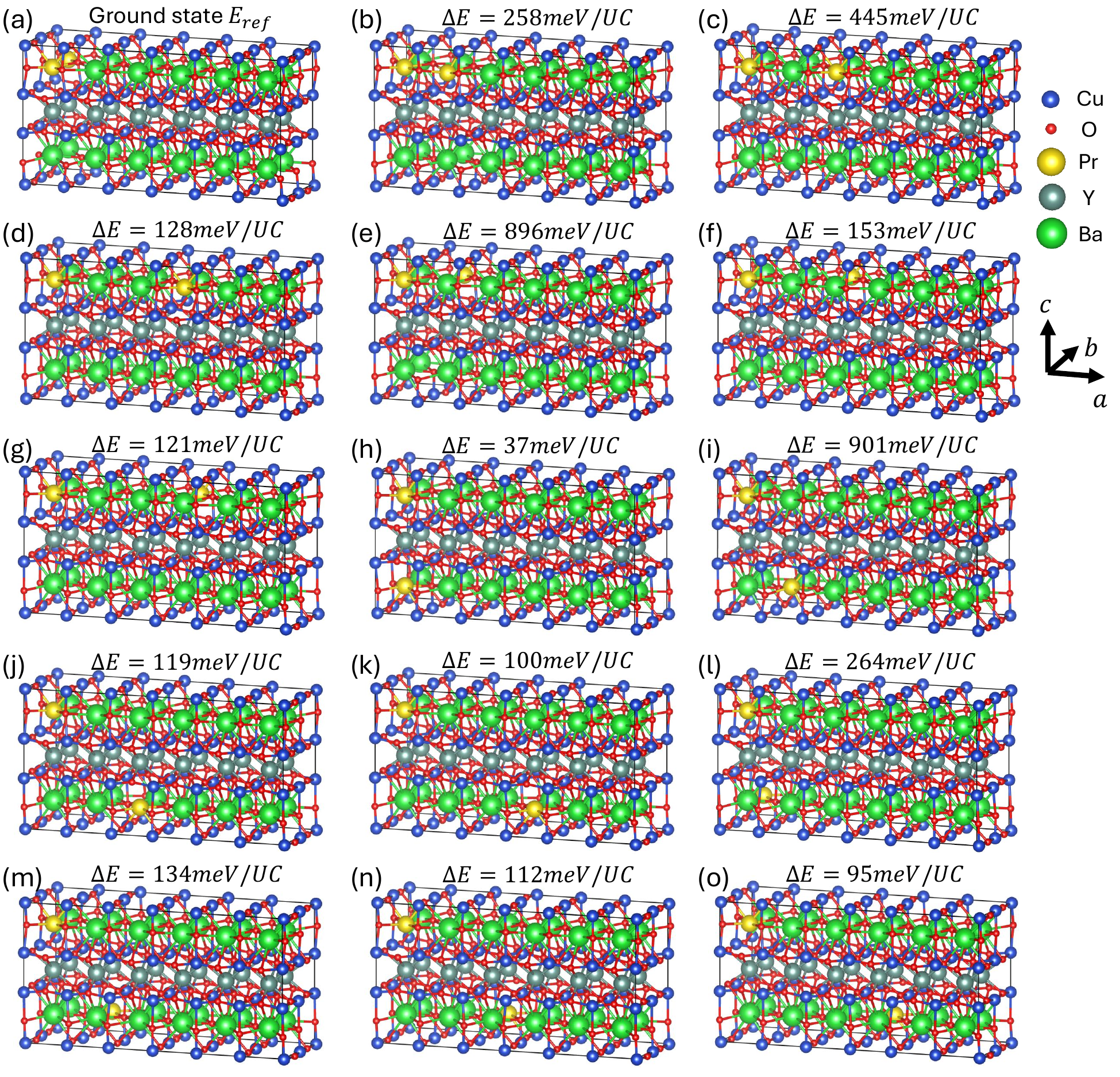}
\end{center}
\caption{  
A set of (meta)stable crystal structures where $1/12$ of the Ba sites are substituted by Pr, corresponding to two Pr atoms per unit cell (outlined by the black boxes). 
(a) Ground-state structure with the lowest total energy. (b-o) Energy penalties $\Delta E$ are reported relative to the reference energy $E_{ref}$ of structure (a) in the unit of meV per unit cell. 
}
\label{fig:app_2Pr}
\end{figure*}

\begin{figure*}[t]
\begin{center}
\includegraphics[scale=0.45]{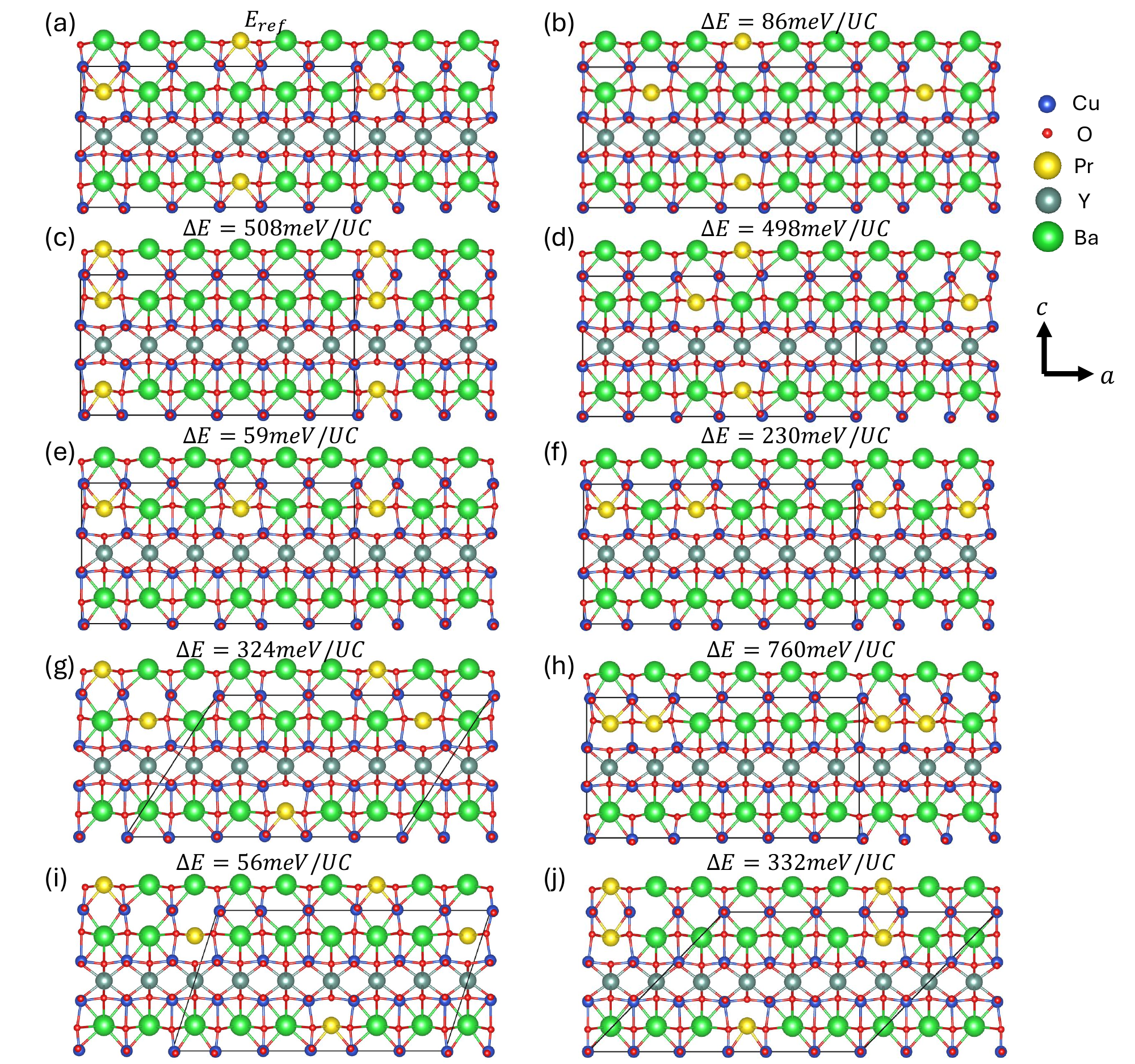}
\end{center}
\caption{  
A set of (meta)stable crystal structures in which $1/6$ of the Ba sites are substituted by Pr, corresponding to four Pr atoms per unit cell (outlined by the black rectangles). Only side views along the $b$ axis are shown, because the alignment along this axis matches that of the lowest-energy structure in Fig.~\ref{fig:app_2Pr}(a). 
(a) Ground-state structure with the lowest total energy. (b-j) Energy penalties $\Delta E$ are reported relative to the reference energy $E_{ref}$ of structure (a) in the unit of meV per unit cell. 
}
\label{fig:app_4Pr}
\end{figure*}
\section{Stable and metastable dopant-driven crystal structures}
\label{app:metastable_crystal}

In this section, we provide a detailed analysis of stable and metastable crystal structures arising from various dopant arrangements. In each calculation, all atoms are constrained to remain nonmagnetic, eliminating any contributions from magnetic ordering within the CuO$_2$ planes and the Pr dopants. Consequently, the energy differences among the configurations presented here result solely from the relative positions of the Pr dopants. This approach allows us to extract dopant–dopant interactions, summarized in Table~\ref{tab:interaction}.

\subsection{Y-site Pr dopants}

Following the experimental Pr content~\cite{ruiz2022stabilization}, substitution of Pr on Y sites replaces approximately $1/3$ of these sites. Under these conditions, the minimal unit cell is a $3\times2\times1$ supercell, containing six Cu atoms per CuO$_2$ plane. Fig.~\ref{fig:app_Y_2Pr}(a) shows the lowest-energy crystal structure obtained from fully relaxed DFT calculations, in which the two Pr dopants align along the $b$-axis. Figs.~\ref{fig:app_Y_2Pr}(b)-(c) illustrate two metastable configurations with higher total energies. The scale of the energy differences induced by varying dopant arrangements is larger than 35 meV per unit cell, which significantly exceeds the energy differences from Cu magnetic ordering shown between Fig.~\ref{fig:DFT_charge_order}(f) and (i) (about 8 meV per unit cell), thereby justifying our treatment of dopant arrangement as a higher-energy-scale phenomenon.

\subsection{Ba-site Pr dopants}

In contrast, when Pr substitution is on the Ba sites, the experimental Pr content only substitutes $1/6$ of the Ba sites, since Ba sites are twice as abundant as Y sites~\cite{ruiz2022stabilization}. Through our DFT calculations, we find that the smallest unit cell capturing the lowest-energy arrangement is a $6\times2\times1$ supercell, containing 12 Cu atoms per CuO$_2$ plane and four Ba sites substituted by Pr atoms.

For clarity, we progressively introduce Pr dopants. Initially, two Ba sites are substituted with Pr, resulting in 15 inequivalent possible arrangements. Fig.~\ref{fig:app_2Pr} summarizes stable and metastable configurations identified from these possibilities. Fully relaxed DFT calculations predict their crystal structures and total energies. The lowest-energy arrangement features Pr dopants aligned along the $b$-axis, shown in Fig.~\ref{fig:app_2Pr}(a), consistent with the alignment observed at the Y sites. Alternative dopant alignments incur significant energy penalties.

Subsequently, we introduce an additional pair of Pr dopants to reach a total of four dopants per unit cell, matching the experimentally observed Pr content. Both dopant pairs follow the alignment established in Fig.~\ref{fig:app_2Pr}(a), oriented along the $b$-axis. We systematically investigate the relative positioning of these two Pr pairs and the corresponding total energies. Fig.~\ref{fig:app_4Pr}(a) illustrates the lowest-energy configuration among these arrangements, where the Pr pairs maximize their separation. By comparing Fig.~\ref{fig:app_4Pr}(a) with Fig.~\ref{fig:app_4Pr}(j), we determine the interaction energy between adjacent bilayers containing Pr pairs along the $c$-axis. Dividing by two (the number of Pr atoms along the $b$-axis), we obtain a 166 meV interaction between two Pr dopants, as listed in Table~\ref{tab:interaction}. Similarly, by analyzing the energy differences among the ten configurations in Fig.~\ref{fig:app_4Pr}, we derive nine effective Pr dopant interactions along the $a$- and $c$-axes, as summarized in Table~\ref{tab:interaction}.

\begin{figure}[t]
\begin{center}
\includegraphics[scale=0.32]{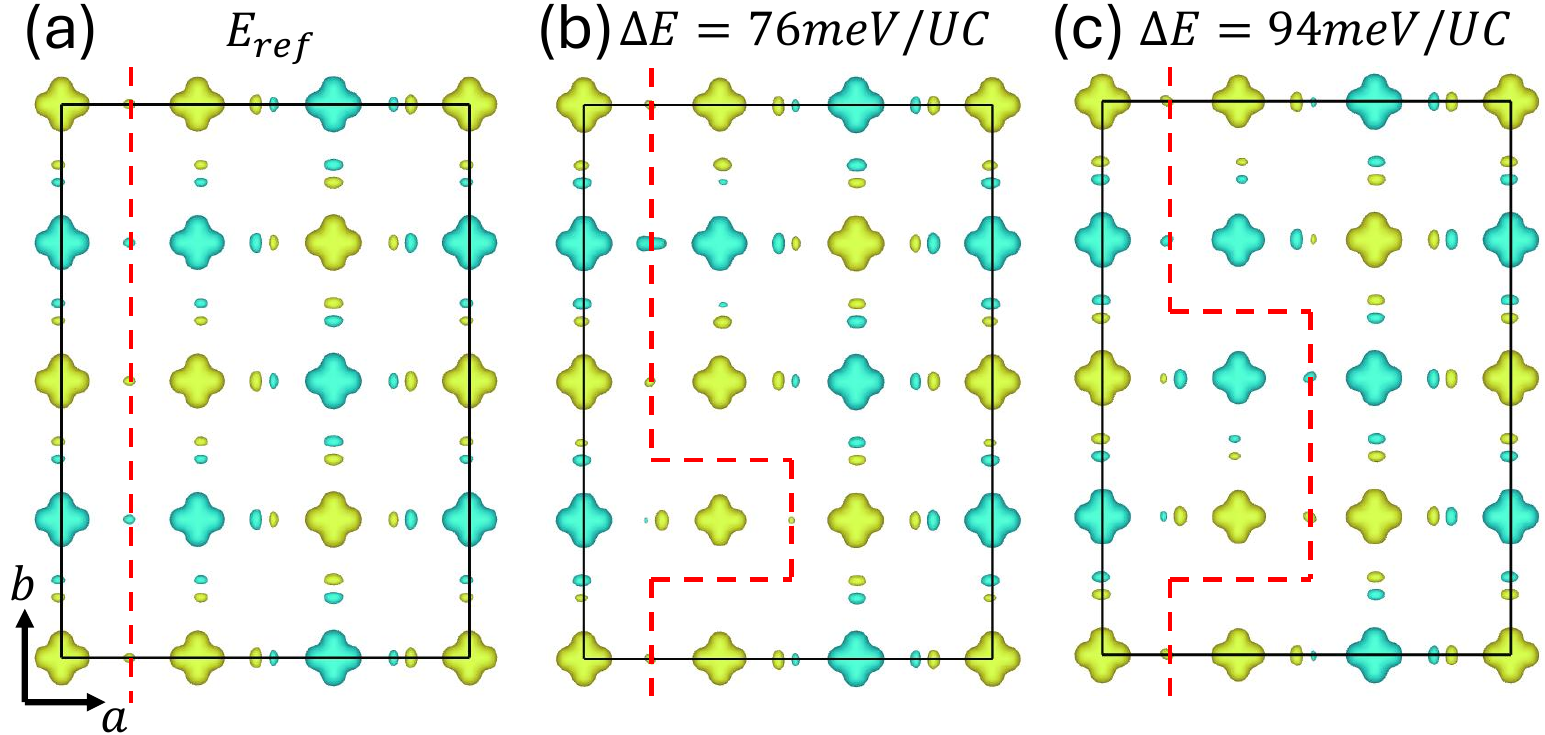}
\end{center}
\caption{  
The top views of a set of (meta)stable magnetic ordered spin structures in pristine YBCO7 with a $3\times4\times1$ supercell (outlined by the black rectangles). 
(a) Spin iso-surface of the ground state, which is equivalent to the stripe order state in Fig.~\ref{fig:DFT_charge_order}(d) with a translational repetition along the $b$-axis. The red dashed line marks out the stripe order domain wall. (b-c) Metastable magnetic order states with tortuous domain walls. Energy penalties $\Delta E$ are reported relative to the reference energy $E_{ref}$ of structure (a) in the unit of meV per unit cell. 
}
\label{fig:app_mag_ab}
\end{figure}

\begin{figure}[t]
\begin{center}
\includegraphics[scale=0.34]{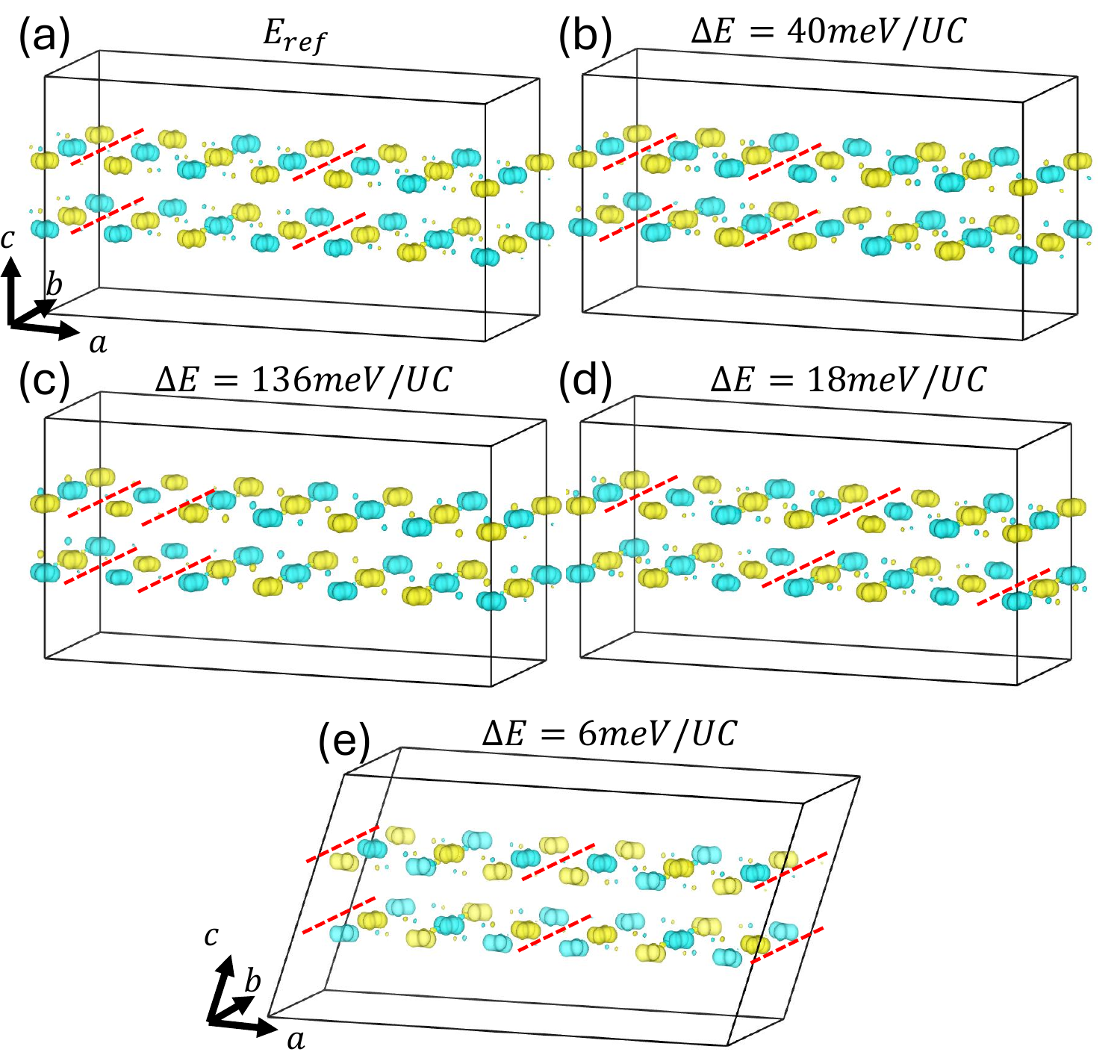}
\end{center}
\caption{  
A set of (meta)stable stripe order structures in pristine YBCO7 with a $6\times2\times1$ supercell (outlined by the black boxes). 
(a) Spin iso-surface of the ground state, which is equivalent to the stripe order state in Fig.~\ref{fig:DFT_charge_order}(d) with a translational repetition along the $b$-axis. (b-e) Metastable stripe order states with the same number of stripe order domain walls. Energy penalties $\Delta E$ are reported relative to the reference energy $E_{ref}$ of structure (a) in the unit of meV per unit cell. 
}
\label{fig:app_mag_pure}
\end{figure}

\begin{figure}[t]
\begin{center}
\includegraphics[scale=0.34]{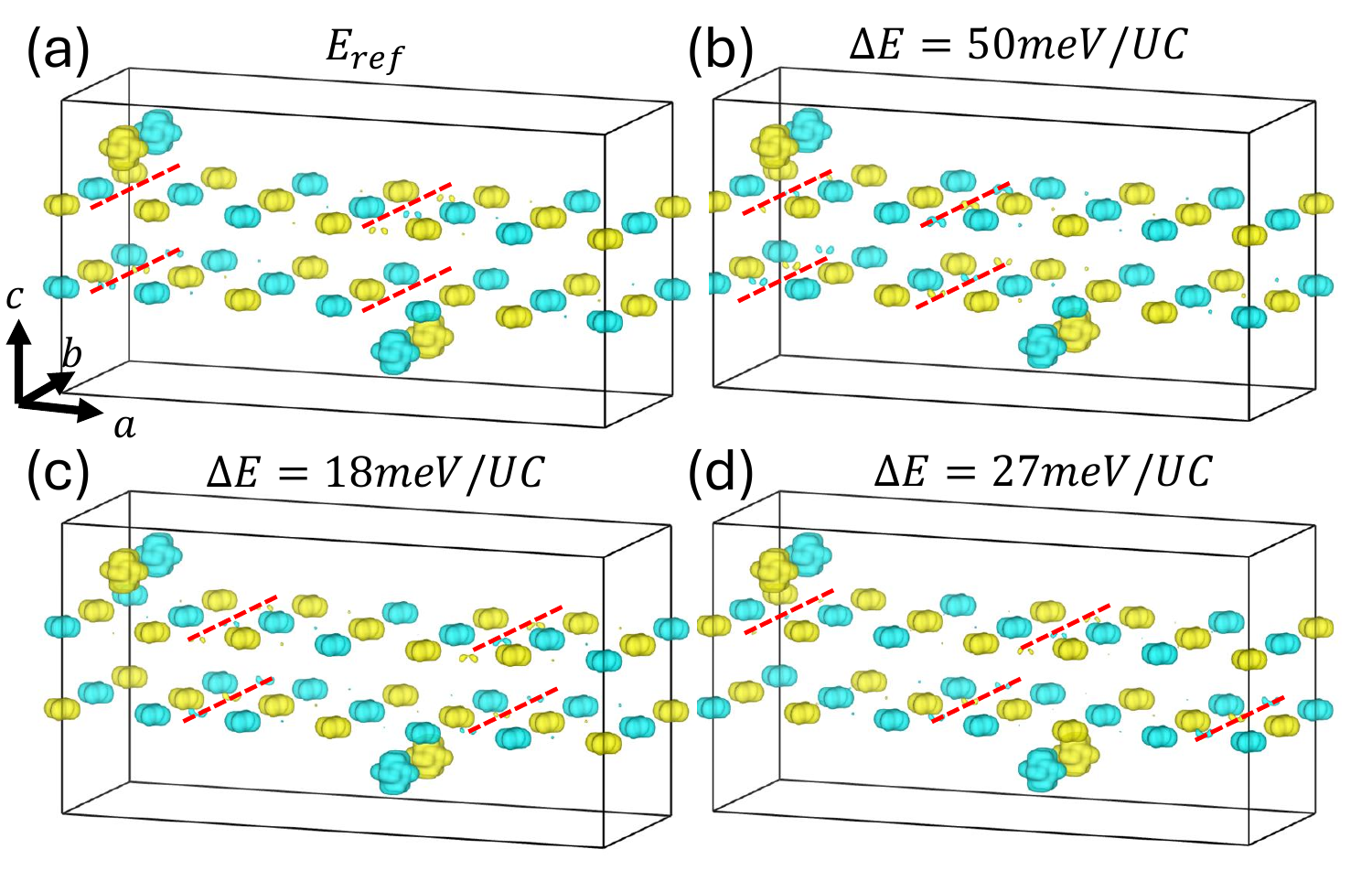}
\end{center}
\caption{  
A set of (meta)stable stripe order structures in Ba-site Pr-doped YBCO7 with a $6\times2\times1$ supercell (outlined by the black boxes). 
(a) Lowest-energy stripe order state, which is the same as Fig.~\ref{fig:DFT_charge_order}(e). (b-d) Metastable stripe order states with the same number of stripe order domain walls. Energy penalties $\Delta E$ are reported relative to the reference energy $E_{ref}$ of structure (a) in the unit of meV per unit cell. 
}
\label{fig:app_mag_Ba}
\end{figure}

\section{Stable and metastable magnetic orders}
\label{app:metastable_mag}

In this section, we provide a detailed analysis of the stable and metastable magnetically ordered states on CuO$_2$ bilayers. In all calculations below, we use the lowest-energy dopant alignments from Appendix~\ref{app:metastable_crystal}, and allow all atoms to develop local moments. Consequently, the energy differences among the configurations presented here result solely from the different magnetic configurations, eliminating the effect of different dopant alignments. For all the calculations, we find that the Pr dopants spontaneously exhibit AFM ordering, consistent with experimental observations~\cite{kebede1989magnetic}, and show a slight energy preference (up to about 2 meV per Pr atom) relative to ferromagnetic (FM) ordering. In the following, we will focus on the magnetic configurations on the CuO$_2$ planes, while leaving Pr dopants (if they exist) to be AFM ordered. 

We first extend the ground state configuration of pristine YBCO7 shown in Fig.~\ref{fig:DFT_charge_order}(d) to a $3\times4\times1$ supercell by replicating the original $3\times2\times1$ supercell along the $b$-axis. This expanded supercell allows exploration of additional possible spin alignments along the $b$-axis beyond the AFM configuration discussed in the main text. Representative stable and metastable states are illustrated in Fig.~\ref{fig:app_mag_ab}. Fig.~\ref{fig:app_mag_ab}(a) corresponds directly to the ground state presented in the main text (Fig.~\ref{fig:DFT_charge_order}(d)), whereas Figs.~\ref{fig:app_mag_ab}(b)-(c) depict metastable states featuring spin misalignments along the $b$-axis, where certain nearest-neighbor spins are parallel; equivalently, one can view these metasable states as having kinks. Such spin misalignments  (kinks) incur energy penalties significantly larger than the energy scales of interest as discussed in the main text, and hence we justifiably assume that the domain walls are straight along the $b$-axis.

Subsequently, all calculations presented below use a $6\times2\times1$ supercell, facilitating various arrangements of CO domain walls within the $ac$ plane and enabling the extraction of effective interactions between these domain walls.

For pristine YBCO7, Fig.~\ref{fig:app_mag_pure}(a) displays the DFT spin isosurfaces of the ground state, highlighting CO domain walls (red dashed lines) spaced by three Cu atoms along the $a$-axis. This configuration represents an expanded version of the state depicted in Fig.~\ref{fig:DFT_charge_order}(d). By repositioning the CO domain walls, we stabilize four representative metastable stripe-ordered states, illustrated in Figs.~\ref{fig:app_mag_pure}(b)-(e). These metastable configurations differ in either their in-plane spacing or their out-of-plane alignment of CO domain walls. For instance, the arrangement in Fig.~\ref{fig:app_mag_pure}(b) incurs an energy penalty of 40 meV per unit cell relative to the reference state in Fig.~\ref{fig:app_mag_pure}(a), indicating a repulsive interaction of approximately 10 meV between two domain walls separated by two Cu atoms along the $a$-axis. Similar analysis yields effective interactions among the CO domain walls summarized in Table~\ref{tab:dw_interaction}.

For Ba-site Pr-doped YBCO7, Fig.~\ref{fig:app_mag_Ba}(a) shows the DFT spin isosurfaces for the ground state, identical to Fig.~\ref{fig:DFT_charge_order}(e) in the main text. Analogously, by repositioning the CO domain walls, we stabilize three representative metastable stripe order states, depicted in Figs.~\ref{fig:app_mag_pure}(b)-(d). Comparing these doped states to their pristine counterparts allows the extraction of dopant-induced effects on CO domain-wall energetics. Our analysis reveals a short-range attraction of approximately 4.5 meV exerted by Pr dopants on adjacent CO domain walls positioned directly above or below. For example, each Pr dopant on the upper Ba layer in Fig.~\ref{fig:app_mag_Ba}(a) has an adjacent CO domain wall below, unlike the configuration in Fig.~\ref{fig:app_mag_Ba}(c). This difference leads to an 18 meV per unit cell energy shift between these states, corresponding to a total contribution from four dopants.

\begin{figure}[t]
\begin{center}
\includegraphics[scale=0.45]{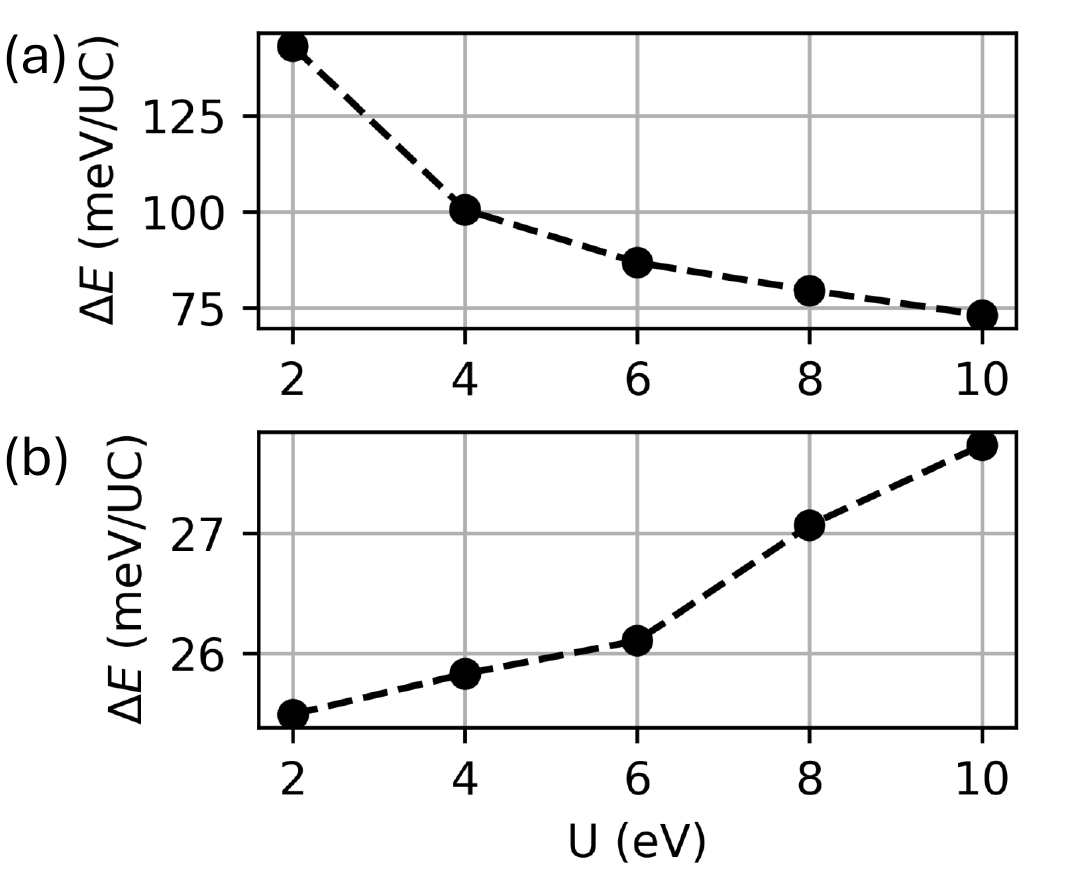}
\end{center}
\caption{  
The energy differences as a function of $U_{\text{Pr}}$ values used in our DFT+$U$ calculations. 
(a) Energy difference $\Delta E$ caused by different dopant alignment, between Fig.~\ref{fig:app_4Pr}(a) and Fig.~\ref{fig:app_4Pr}(b). 
(b) Energy difference $\Delta E$ caused by different CO domain wall alignment, between Fig.~\ref{fig:app_mag_Ba}(a) and Fig.~\ref{fig:app_mag_Ba}(d). 
}
\label{fig:app_PrU}
\end{figure}

\section{Choice of $U$-values}
\label{app:dftu}

The DFT calculations possess several well-known limitations~\cite{cohen2008insights}. For instance, the local-density approximation (LDA) typically underestimates band gaps by approximately 40\%, particularly in semiconductors and insulators~\cite{perdew1983physical}. This underestimation arises primarily due to self-interaction errors present in the approximate exchange-correlation functionals~\cite{perdew1981self}. To address this, the DFT+$U$ approach applies a Hubbard correction parameter, $U$, selectively to specific states—usually $d$- or $f$-shell orbitals~\cite{anisimov1991band}. This approach markedly improves predictions of band gaps and magnetic ordering, particularly for transition metal oxides~\cite{kirchner2021extensive}.

Specifically for Cu $d$-orbitals, prior research has identified an optimal Hubbard parameter $U_{\text{Cu}}=4$ eV for copper oxides, which reproduces experimental magnetic moments and band gaps accurately~\cite{wang2006oxidation}. This $U_{\text{Cu}}$ value has been consistently employed in previous theoretical studies of copper-based superconductors~\cite{yelpo2021electronic, deng2019higher, jin2024first, jin2025interlayer}. Adopting this value, our DFT calculations yield an AFM magnetic moment of $0.56 \mu_B$ on planar Cu atoms in the parent compound YBa$_2$Cu$_3$O$_6$ (YBCO6), aligning well with the experimentally measured local moments ranging from $0.52$ to $0.60\mu_B$\cite{burlet1988neutron, shamoto1993neutron}. Additionally, by comparing energies between AFM and FM configurations, we determine a magnetic exchange interaction strength of $J=134$ meV, which also closely matches experimental findings \cite{shamoto1993neutron, hayden1996high, le2011intense}. Given that these observables are tightly linked to the selected $U_{\text{Cu}}$, there is limited flexibility in adjusting this parameter.

Regarding Pr $f$-orbitals, previous DFT+$U$ studies have not found a unique optimal $U_{\text{Pr}}$, yet values typically fall within the range of 6 to 10 eV \cite{liechtenstein1995quantitative, ghanbarian2006effects, ghanbarian2008different}. Figure \ref{fig:app_PrU} illustrates the variation of representative energy differences between stable and metastable states across a range of $U_{\text{Pr}}$ from 2 to 10 eV. Specifically, Fig.~\ref{fig:app_PrU}(a) demonstrates that the energy difference between two distinct dopant alignments varies by roughly 13\% when $U_{\text{Pr}}$ ranges between 6 and 10 eV. Similarly, the energy difference for two different domain-wall alignments varies by about 7\% within the same $U_{\text{Pr}}$ interval. Consequently, the choice of $U_{\text{Pr}}$ within this range minimally impacts the energy models discussed in the main text.

\begin{figure}[t]
\begin{center}
\includegraphics[scale=0.37]{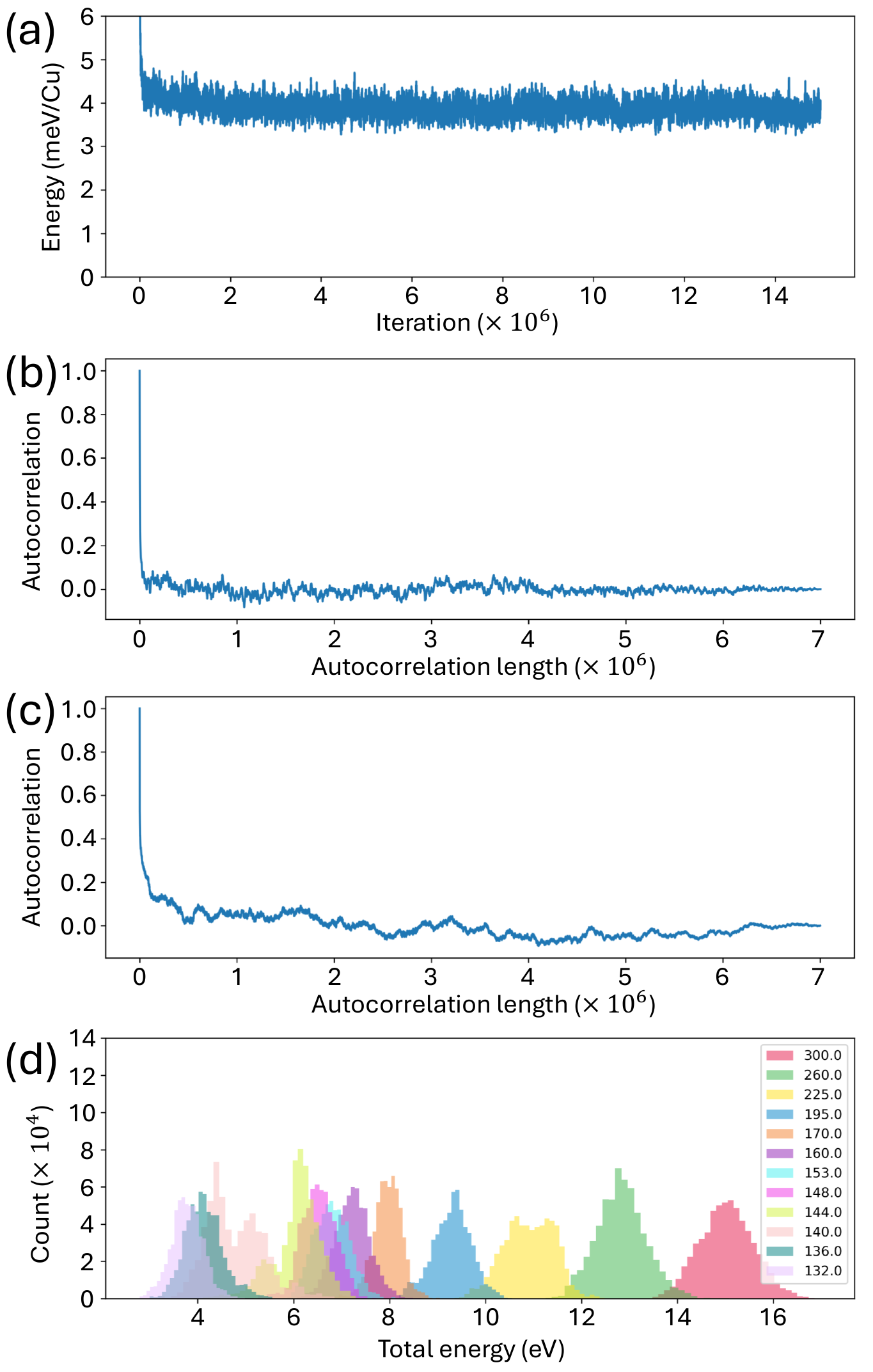}
\end{center}
\caption{  
(a) The total energy per Cu atom as a function of Monte Carlo iterations. The data comes from the 300K simulation of the dopant alignment. All calculations have 15 million iterations. (b) Normalized energy autocorrelation computed from the last 7 million iterations in (a). (c) Normalized energy autocorrelation computed from the last 7 million iterations in the 136K simulation of the dopant alignment. (d) Total energy histogram of the last 4 million iterations for each of the different temperatures in our parallel tempering. The color legend lists temperatures in units of Kelvin. 
}
\label{fig:app_equil}
\end{figure}

\section{Monte Carlo simulation details}
\label{app:mc_details}

In this section, we provide technical details related to our Monte Carlo simulations. To confirm equilibrium, we first examine the relationship between total energy and the number of iterations. As an illustrative example, Fig.~\ref{fig:app_equil}(a) presents the total energy as a function of Monte Carlo iterations for a relatively high-temperature (300~K) calculation of dopant alignment. The total energy fluctuates around a steady value after approximately two million iterations. Equilibrium is further validated by computing the energy autocorrelation for the final 7 million iterations, as shown in Fig.~\ref{fig:app_equil}(b). The energy autocorrelation rapidly decreases as the autocorrelation length increases. Additionally, Fig.~\ref{fig:app_equil}(c) displays the energy autocorrelation from a lower-temperature simulation, confirming a similarly rapid decay and fluctuations around zero. Consequently, we are assured that equilibrium is established over the last 5 million iterations in all simulations, allowing reliable estimation of equilibrium properties with reasonable statistical uncertainties, as demonstrated in the main text.

To mitigate potential trapping in metastable states during low-temperature simulations, we employ parallel tempering Monte Carlo. This approach leverages high-temperature simulations to explore both stable and metastable configurations, facilitating escape from local minima in lower-temperature runs. Fig.~\ref{fig:app_equil}(d) illustrates the distributions of total energy at various temperatures used in our parallel tempering, highlighting sufficient overlap between adjacent temperatures, a crucial condition for effective parallel tempering.

\section{Diffusion rate}
\label{app:diffusion}

While the diffusion coefficient of Pr in YBCO7 has not been measured directly, insights into dopant dynamics can be gained from diffusion studies of a typical dopant, such as Ag in YBCO7. Experimental studies \cite{gorur2005effect} have reported a diffusion coefficient for high temperatures with a thermally activated (Arrhenius) form:
\begin{equation}
D = 1.9\times10^{-6}\exp(-0.73\mbox{ eV}/k_BT)\ \mbox{cm}^2\mbox{s}^{-1}\,.
\end{equation}
At high temperatures, dopants diffuse freely and their spatial locations follow the Boltzmann distribution. As temperature decreases, diffusion slows dramatically, raising the practical question: below what temperature is the dopant configuration effectively static over a typical experimental timescale (e.g., one day)?

Assuming dopant motion occurs via elementary hops over a distance of approximately 3~\AA, the characteristic hopping rate is given by:
\begin{equation}
r = 1.3 \times 10^{10} \exp\left(-\frac{0.73\ \mathrm{eV}}{k_B T}\right)\ \mathrm{s}^{-1}.
\end{equation}
The dopant configuration is effectively frozen when the average hopping time $1/r$ exceeds the duration of an experiment. Using this criterion, setting $1/r$ equal to one day yields a characteristic ``freezing'' temperature of approximately $T_F=244$~K. Below this temperature, dopant rearrangement occurs on timescales much longer than typical experiments, and the configuration can be considered static. Moreover, cooling the system faster than roughly 1 K/day from around $T_F$ suppresses equilibration, effectively trapping the system in a metastable dopant configuration reflective of higher-temperature states.

\begin{table}
\begin{tabular}{c|cccc}
Element & $D_0$ (cm$^2$/s) & $E_b$ (eV) & $T_F$ (K) & ref. \\
\hline
Ag & $1.9\times10^{-6}$ & 0.73 & 244 & \cite{gorur2005effect} \\
Ag & $1.0\times10^{-2}$ & 1.1 & 157 & \cite{kishio1989oxygen} \\
Au & $3.2\times10^{-4}$ & 1.0 & 292 & \cite{routbort1994oxygen} \\
Au & $3.4\times10^{-6}$ & 0.85 & 196 & \cite{baetzold1990atomistic} \\
In & $1.1\times10^{-8}$ & 0.26 & 103 & \cite{o1988study} \\
Sn & $8.0\times10^{-6}$ & 0.58 & 203 & \cite{o1988study} \\
Ni & $1.0\times10^{-2}$ & 1.3 & 186 & \cite{o1988study} \\
Ni & $3.2\times10^{-10}$ & 0.17 & 278 & \cite{o1988study} \\
Fe & $1.5\times10^{-4}$ & 1.25 & 221 & \cite{maier1993defect} \\
Fe & $1.8\times10^{-6}$ & 0.90 & 204 & \cite{maier1993defect} \\
\end{tabular}
\caption{Diffusion coefficients $D=D_0\exp(-E_b/kT)$ and effective freezing temperatures $T_F$ (estimated with 1 K/day cooling rate) of different dopant elements in bulk YBCO7. }
\label{tab:frozen}
\end{table}

Similarly, we compute the effective freezing temperature for other dopants using the same approach. Table~\ref{tab:frozen} lists several representative dopants along with their measured diffusion coefficients. In all cases, the freezing temperatures $T_F$ corresponding to a cooling rate of 1~K/day fall within the range of 103~K to 292~K. This supports the expectation that the Pr dopant likewise encounters a freezing temperature $T_F$ during the experimental cooling from room temperature to 50~K.

\bibliography{v1}
\end{document}